\documentclass[11pt]{article}

\usepackage[final]{acl}

\usepackage{times}
\usepackage{latexsym}

\usepackage[T1]{fontenc}

\usepackage[utf8]{inputenc}

\usepackage{microtype}

\usepackage{inconsolata}
\usepackage{hyperref}

\usepackage{graphicx}
\usepackage{tikz}
\usepackage{tikzpagenodes}
\usepackage{subcaption}
\usepackage{amsmath}
\usepackage{amssymb}
\usepackage{multirow}

\usepackage{longtable}
\usepackage{xcolor}
\usepackage{booktabs}

\definecolor{ab_color}{HTML}{0077b6}

\definecolor{rr_color}{HTML}{2d6a4f}

\definecolor{iz_color}{HTML}{e76f51}

\definecolor{ja_color}{HTML}{7b2d8b}

\definecolor{sr_color}{HTML}{c0392b}

\title{Closing the Operational Gap in Semantic Caching\thanks{Accepted at EMNLP 2026 (Industry Track).}}

\author{
  \textbf{Aditeya Baral\thanks{Equal contribution.}\thanks{Work done as an Intern at Redis.}},
  \textbf{Radoslav Ralev\footnotemark[2]},
  \textbf{Iliya Sotirov Zhechev},
  \\
  \textbf{Srijith Rajamohan},
  \textbf{Jen Agarwal}
  \\
  New York University
  \quad
  Redis
  \\
  {
  \texttt{aditeyabaral@nyu.edu} \quad
  \texttt{\{firstname.lastname\}@redis.com}
  }
  \\[3pt]
  \href{https://github.com/aditeyabaral/operational-gap-semantic-caching}{\nolinkurl{github.com/aditeyabaral/operational-gap-semantic-caching}}
}

\begin{document}
\maketitle

\begin{tikzpicture}[remember picture, overlay]
\node[anchor=north east, yshift=1.5cm]
  at (current page text area.north east)
  {\includegraphics[height=1cm]{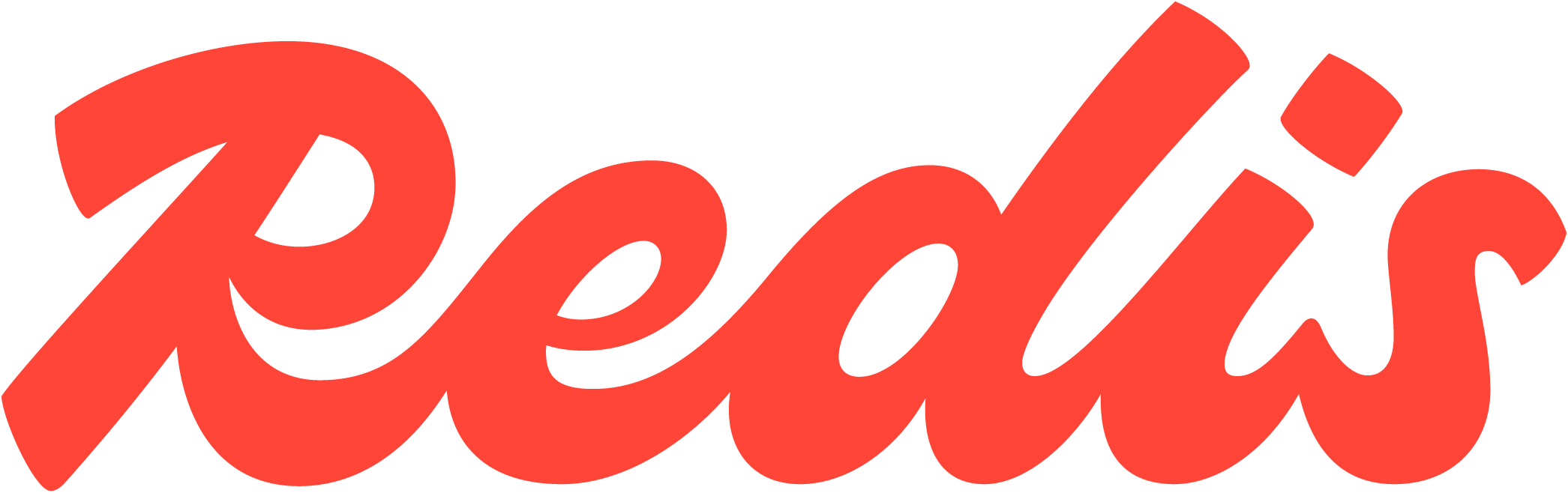}};
\end{tikzpicture}

\begin{abstract}
Semantic caching cuts LLM inference costs by serving a cached response to semantically similar queries. Standard practice evaluates these systems using PR-AUC, a metric that only measures how well scores rank and ignores whether they are usable at a fixed threshold. We show this mismatch leads to systematically poor deployment choices, as models with the highest PR-AUC are often the worst in operation. We introduce Precision--Cache Hit Ratio (P-CHR) AUC, a cache-aware metric that measures precision across cache utilization levels, and Operational Retention Rate (ORR), which captures how much offline ranking quality survives at deployment. We decompose the operational gap between offline and deployed quality into a recoverable threshold-utility component and an irreducible structural component fixed by the dataset's positive rate. Our experiments show that the threshold-utility gap is governed by the training objective rather than data scale, and yields only to re-normalizing scores over the candidate pool or changing the training objective. Ultimately, model selection for semantic caching is a threshold-utility problem, not a ranking one, and measuring it is the first step to closing the gap.

\end{abstract}

\section{Introduction}

Large language models (LLMs) have become integral to modern applications, powering conversational AI, code generation, and knowledge retrieval systems. However, their operational costs remain prohibitively high due to expensive computational infrastructure including high-end GPUs, distributed serving systems, and substantial memory resources~\cite{frantar2023gptqaccurateposttrainingquantization}. Commercial LLM APIs compound these costs through token-based pricing and rate limits that constrain throughput. Studies indicate a significant fraction of these expensive inference requests are redundant: approximately 33\% of online web searches~\cite{10.1145/775152.775156, xie2002locality, MARKATOS2001137} and LLM chat logs~\cite{gill2025meancache} are repeated or semantically similar, presenting a substantial optimization opportunity through \textit{semantic caching}~\cite{bang-2023-gptcache, gill2025meancache}.

Unlike traditional key-value caches requiring exact string matching, semantic caches leverage dense embeddings~\cite{reimers2019sentencebertsentenceembeddingsusing, karpukhin2020densepassageretrievalopendomain} to recognize duplicate, paraphrased, or contextually equivalent queries. When the similarity score between a query and a cached response exceeds a predefined threshold $\tau$, the system retrieves the cached LLM response rather than invoking the LLM again. For instance, ``how do I reset my password?'' and ``forgot login credentials'' should retrieve the same cached response despite differing in wording. To further improve precision, a cross-encoder reranker~\cite{nogueira2020passagererankingbert} can jointly encode each query--candidate pair and assign a refined relevance score, replacing the retriever score for the threshold decision. Prior work has demonstrated that fine-tuned, domain-specific embedding models can achieve state-of-the-art performance in semantic caching while maintaining computational efficiency~\cite{gill2025advancingsemanticcachingllms}. 

A key property of this deployment setting is that the threshold $\tau$ transforms a similarity score into a binary cache decision: scores above $\tau$ fire the cache; scores below it fall through to the LLM. The deployment objective of a semantic cache is therefore to maintain high precision, i.e., the fraction of correct cache hits across a range of cache utilization levels as $\tau$ varies. A model with well-placed scores can be operated at any practical $\tau$ and reliably return correct cached responses; a poorly-placed model may rank positives above negatives yet produce precision collapse the moment $\tau$ is set to a value where the cache fires at a useful rate. Prior semantic caching evaluations assess decision quality via Precision-Recall AUC (PR-AUC)~\cite{gill2025advancingsemanticcachingllms, gill2025meancache}, a metric agnostic to score magnitudes because it only measures relative ordering. We show that this mismatch between the evaluation metric and the deployment objective leads to systematic errors in model selection, across both retriever-only and retrieval+reranking systems.

To address this mismatch, we propose two new cache-aware metrics: \textbf{P-CHR AUC} (Precision--Cache Hit Ratio AUC), which captures the precision--utilization tradeoff at every threshold, and the \textbf{Operational Retention Rate (ORR)}, which measures how much of a model's offline ranking quality survives at deployment. Together they measure what PR-AUC ignores. We define the \textit{operational gap} as the difference between a system's PR-AUC and its P-CHR AUC, capturing the divergence between offline ranking quality and threshold-based deployment performance. This gap further decomposes into two components: a \textit{structural gap}, which is irreducible and fixed by the dataset's positive rate, and a \textit{threshold-utility gap}, a recoverable loss arising from where a model places its scores rather than from probability miscalibration.


\paragraph{Contributions.} We make the following contributions:

\begin{enumerate}
  \item \textbf{New metrics.} \textbf{P-CHR AUC}, a cache-aware metric that captures the precision--utilization tradeoff at deployment thresholds, and the \textbf{Operational Retention Rate (ORR)}, the share of offline ranking quality a model retains at deployment (\S\ref{sec:metrics}).

  \item \textbf{Operational gap decomposition.} A decomposition of the operational gap into an irreducible structural component and a recoverable threshold-utility component (\S\ref{sec:metrics}).

  \item \textbf{Empirical study.} Evaluation of 9 retrievers and 10 rerankers on 74,265 test queries across both offline and deployment settings (\S\ref{sec:results}).

  \item \textbf{Dataset and model releases.} \textit{LangCache SentencePairs}\footnote{\url{https://huggingface.co/collections/redis/langcache-datasets}}, large-scale sentence-pair datasets (40M training pairs) and the \textit{LangCache-Embed-v3}\footnote{\url{https://huggingface.co/redis/langcache-embed-v3-small}} and \textit{LangCache Reranker}\footnote{\url{https://huggingface.co/collections/redis/langcache-re-ranker-v2}; v1 collection also
  available} models, all curated for semantic caching.

\end{enumerate}

\section{Methodology}
\label{sec:methodology}

\subsection{Dataset Curation}
\label{sec:dataset}

Training retrieval and reranking models for semantic caching requires large-scale sentence-pair datasets with high linguistic diversity and semantic complexity. Existing benchmarks are typically small by modern standards (often fewer than 1M pairs) and lack the breadth required for robust generalization across tasks from technical documentation to customer support. 

We curated the \textbf{LangCache SentencePairs} collection by aggregating sentence-pair data from diverse sources spanning paraphrase detection, semantic textual similarity, question answering, adversarial examples, domain-specific terminology, and others to ensure broad coverage of real-world LLM usage. Each example is a tuple $(q, c, y)$ where $q$ is the query, $c$ is the candidate, and $y \in \{0, 1\}$ denotes semantic equivalence. Training data class balance varies with version, with well-defined validation and test splits across all versions.

We release three progressively scaled versions (Table~\ref{tab:dataset-stats}). \textbf{Version 1} (1.05M pairs) establishes a high-quality baseline with maximal linguistic variety. \textbf{Version 2} (8M pairs) augments v1 with a synthetic paraphrase generation pipeline~\cite{gill2025advancingsemanticcachingllms}, increasing training signal while maintaining the same validation split. \textbf{Version 3} (40M pairs) incorporates large-scale back-translated and synthetically generated paraphrase corpora, a 38$\times$ scale increase over v1. All evaluations in this work are conducted on the v3 test split (74,265 pairs, 45\% positive). Full dataset details are in Appendix~\ref{sec:dataset-curation}.

\subsection{Model Training}
\label{sec:training}

\paragraph{LangCache-Embed-v3.} We fine-tune \textbf{LangCache-Embed-v3}, a domain-specific bi-encoder initialized from \textbf{all-MiniLM-L6-v2}~\cite{wang2020minilm, reimers2019sentencebertsentenceembeddingsusing}, a 22.6M-parameter MiniLM encoder. We train it on LangCache SentencePairs v2 (8M pairs) with an in-batch contrastive objective, the ArcFace additive angular margin loss~\cite{deng2019arcface}. The model produces unit-normalized embeddings for cosine-similarity comparison. Full training details are in Appendix~\ref{sec:training-config}.

\paragraph{LangCache Rerankers.}
We fine-tune \textbf{LangCache Reranker} models, domain-specific cross-encoders initialized from \textbf{GTE-Reranker-ModernBERT-base}~\cite{zhang-etal-2024-mgte, li2023generaltextembeddingsmultistage, warner-etal-2025-smarter}, a 150M parameter ModernBERT encoder. We compare two training objectives that differ fundamentally in what their scores represent. \textbf{Binary Cross-Entropy (BCE)} treats reranking as binary classification, minimizing cross-entropy loss against hard relevance labels. \textbf{Multiple Negatives Ranking Loss (MNRL)}~\cite{reimers2019sentencebertsentenceembeddingsusing} is a contrastive objective that learns relative ordering instead of absolute probability boundaries. 

We train two model generations: \textbf{LangCache Rerankers v1} on LangCache SentencePairs v1 (1M pairs) and \textbf{LangCache Rerankers v2} on LangCache SentencePairs v3 (40M pairs), each with BCE and MNRL variants. Reranker versions reflect training generations rather than dataset versions. We skip the intermediate v2 dataset by design: its 8$\times$ scale increase over v1 is insufficient to isolate scale effects cleanly, whereas the 38$\times$ jump to v3 provides the contrast needed to separate the contributions of training objective and scale. Full training details are in Appendix~\ref{sec:training-config}.

\section{Metrics}
\label{sec:metrics}

Each query $q$ has an associated ground-truth candidate $c^*$ and a binary label $y \in \{0, 1\}$ indicating whether $c^*$ is a genuine semantic match. A system assigns a score $s(q, c^*)$ to the ground-truth candidate, used as a ranking signal for offline evaluation. In deployment, the cache fires for threshold $\tau$ when the top-ranked retrieved candidate $\hat{c}(q)$ has score $\hat{s}(q) = \max_k s(q, c_k) \geq \tau$. Let $N$ denote the total number of queries and $p = |\{q : y_q = 1\}| / N$ the positive rate.

\paragraph{Four notions of score quality.} Our metrics concern how scores behave at a fixed decision threshold, which we keep distinct from probability calibration. We separate four properties. \emph{Probability calibration} asks whether a score equals $P(\text{match})$, measured by ECE or NLL (Appendix~\ref{sec:distribution-plots}). \emph{Score placement} is where a model's scores sit on $[0,1]$ relative to the threshold. \emph{Threshold utility} is the precision attainable across cache utilization levels, which is what P-CHR AUC measures. \emph{Deployment robustness} is how wide a threshold band holds a target precision (Appendix~\ref{sec:extended-discussion}). This paper measures the latter two: a score can be poorly calibrated as a probability yet have high threshold utility, and vice versa.

\subsection{Classification Metrics}
\label{sec:metrics-classification}

We report \textbf{PR-AUC}, the area under the Precision--Recall curve traced as the threshold $\tau$ is swept, where precision and recall threshold the ground-truth score $s(q,c^*)$. PR-AUC summarizes ranking quality without committing to a threshold. The confusion-matrix counts, along with the definitions of precision and recall for thresholded decisions, are in Appendix~\ref{sec:metric-derivations}.

\subsection{Cache-Aware Metrics}
\label{sec:metrics-cache}

Standard classification metrics evaluate whether $s(q, c^*)$ exceeds a threshold, an offline proxy that does not model what the cache actually serves. In deployment, the cache fires on the top-ranked candidate $\hat{c}(q)$ from ANN retrieval, which may differ from $c^*$ even when $c^*$ is in the retrieved pool. We introduce metrics that capture this deployment behavior.

A \textbf{cache fire} occurs when $\hat{s}(q) \geq \tau$. A fire is \textbf{valid} when $\hat{c}(q) = c^*$ and $y_q = 1$: the cache serves the correct candidate to a genuine positive query.

\paragraph{Cache Hit Ratio (CHR).}

\begin{equation}
  \text{CHR}(\tau) = \frac{|\{q : \hat{s}(q) \geq \tau\}|}{N}
\end{equation}

CHR measures the fraction of all queries for which the cache fires, regardless of correctness. It is the primary driver of cost savings. Appendix~\ref{sec:chr-vs-recall} discusses why CHR is the appropriate operationally measurable quantity for production caches.

\paragraph{Deployment Precision.} The cache-decision precision $\text{Precision}(\tau)$ is the fraction of cache fires that are valid, computed on the top-1 score $\hat{s}(q)$. It is distinct from the classification precision of \S\ref{sec:metrics-classification}, which thresholds the ground-truth score $s(q, c^*)$. We formalize it as the ratio of valid fires to all fires (VCHR/CHR) in Appendix~\ref{sec:metric-derivations}.

\paragraph{P-CHR AUC.}

Precision-CHR curves plot Precision$(\tau)$ against CHR$(\tau)$ as $\tau$ is swept. As $\tau$ decreases, CHR rises (the cache serves more queries) but precision typically falls (more false positives enter). The area under this curve is:
\begin{equation}
  \text{P-CHR AUC} = \int_{0}^{1} \text{Precision}\bigl(\text{CHR}^{-1}(c)\bigr)\, dc
\end{equation}

\noindent P-CHR AUC quantifies how well a model maintains precision as cache utilization grows: a model with high P-CHR AUC offers many viable operating points where both precision and cache hit ratio are acceptable.

\noindent Unlike PR-AUC, which ranks the offline ground-truth score $s(q,c^*)$, P-CHR AUC scores the deployed top-1 decision $\hat{s}(q)$: a model can rank $c^*$ well offline (high PR-AUC) yet place its firing scores so that precision collapses at any practical CHR. We compute P-CHR AUC \emph{exactly}, integrating over every distinct score rather than approximating the area on a fixed threshold grid. It is invariant to any strictly monotone rescaling of scores, so it is directly comparable across models with different score semantics (ColBERT softmax, BCE probabilities, or MNRL logits) and, as a corollary, cannot be improved by post-hoc monotone calibration.

\subsection{Operational Gap Decomposition}
\label{sec:metrics-gap}

For any model that ranks better than chance, P-CHR AUC is strictly lower than PR-AUC, even for a perfect ranker, because the x-axes of the two curves differ. The \textbf{operational gap} measures this total divergence:
\begin{equation}
  \Delta_{\text{op}} = \text{PR-AUC} - \text{P-CHR AUC}
\end{equation}

\noindent Part of this gap is \textit{irreducible}: even a perfect ranker (all positives scoring above all negatives) caps P-CHR AUC at $p(1 - \ln p) \approx 0.809$ ($p \approx 0.45$ in our test set, derivation in Appendix~\ref{sec:metric-derivations}), so the \textbf{structural gap}
\begin{equation}
  \Delta_{\text{str}} = 1 - p(1 - \ln p)
\end{equation}
\noindent is irreducible: no post-processing can eliminate it on a dataset with positive rate $p$. The remaining \textbf{threshold-utility gap}

\begin{equation}
  \Delta_{\text{util}} = \max\!\bigl(0,\; \Delta_{\text{op}} - \Delta_{\text{str}}\bigr)
\end{equation}

\noindent reflects recoverable mis-placement of scores. When $\Delta_{\text{op}} > \Delta_{\text{str}}$, score compression or boundary collapse depresses P-CHR AUC below what a well-placed model of equal ranking quality would achieve. When $\Delta_{\text{op}} \leq \Delta_{\text{str}}$, the operational gap falls within the structural floor: the model's score placement is already as well-suited to threshold-based decisions as the metric structure permits, and $\Delta_{\text{util}} = 0$ by definition. Because P-CHR AUC is invariant to monotone rescaling, post-hoc calibration leaves $\Delta_{\text{util}}$ unchanged (Appendix~\ref{sec:calibration-fitting}); it is reduced instead by re-normalizing scores over the candidate pool (Appendix~\ref{sec:normalization-ablation}) or by the training objective.

We further define the \textbf{Operational Retention Rate (ORR)} as the fraction of offline ranking quality retained in deployment:
\begin{equation}
  \text{ORR} = \frac{\text{P-CHR AUC}}{\text{PR-AUC}} \in (0, 1]
\end{equation}
\noindent For a perfectly-placed model with PR-AUC $= 1$, ORR reaches the structural ceiling $p(1 - \ln p)$. ORR complements $\Delta_{\text{util}}$: while $\Delta_{\text{util}}$ measures the absolute recoverable gap, ORR captures how efficiently a model converts offline ranking quality into deployment precision.

\section{Experimental Setup}
\label{sec:setup}

\subsection{Test Set}

We evaluate on the LangCache SentencePairs v3 test split: 74,265 sentence pairs with 45\% positive labels ($p \approx 0.45$), yielding a structural gap of $\Delta_{\text{str}} \approx 0.19$. This split was held out during all model training. All reported metrics are computed on this split.

\subsection{Models}

\paragraph{Retrievers.} We evaluate 9 bi-encoder embedding models as stage-1 retrievers: 3 domain-specific models fine-tuned for semantic caching and 6 general-purpose embedding models. All produce unit-normalized embeddings for cosine similarity comparison. Full model names and citations are listed in Table~\ref{tab:retriever-baselines}.

\paragraph{Rerankers.} We evaluate 10 reranking models across three families: our \textbf{LangCache Rerankers} (4 models, with both BCE and MNRL variants at two training scales as described in \S\ref{sec:training}); general-purpose cross-encoders (2 models); and multi-vector models from the ColBERT-family (4 models), which produce token-level representations and score via MaxSim aggregation. Full model names are listed in Table~\ref{tab:calibration-gap}.

\subsection{Evaluation Pipeline}

A Redis semantic cache\footnote{\url{https://docs.redisvl.com/en/latest/api/cache.html\#semanticcache}} is populated with dense embeddings of all unique candidate sentences in the test split. For each query $q$, we run a two-stage pipeline: (1) \textbf{$K$-NN Retrieval}: retrieve the top-$K$ candidates by exact cosine similarity search over the candidate pool;\footnote{Production caches typically approximate this step with an ANN index such as HNSW~\cite{johnson2017billionscalesimilaritysearchgpus, malkov2018efficientrobustapproximatenearest}; we use exact search for reproducibility.} (2) \textbf{Reranking}: the cross-encoder (or multi-vector model) scores all $K$ retrieved query--candidate pairs, and the highest-scoring candidate is selected as the predicted cache match. We fix $K = 50$ throughout since it reflects the larger candidate pools typical of production caches. Appendix~\ref{sec:k-sensitivity} examines sensitivity to the candidate pool size.

For score normalization, BCE reranker outputs are treated as direct probabilities $\sigma(z)$; MNRL outputs are mapped to $[0,1]$ via sigmoid applied to the raw logit. ColBERT-family outputs are unnormalized MaxSim aggregates; they are softmax-normalized across the $K$ retrieved candidates, mapping $\hat{s}(q)$ to $[0,1]$. 

For any ground-truth positive $c^*$ not retrieved in the top-$K$, we assign $s(q, c^*) = 0.0$. This imposes the retriever recall ceiling on all downstream metrics: a miss at retrieval cannot be recovered by reranking. We compute P-CHR and P-VCHR AUC by sweeping $\tau$ through every distinct score so the reported area is the exact value of the integral.

\section{Results}
\label{sec:results}

\subsection{Retriever Baselines}
\label{sec:results-retriever}

Table~\ref{tab:retriever-baselines} reports PR-AUC and P-CHR AUC for the 9 retrievers at $K=50$. Domain-specific models lead offline and substantially outperform general-purpose models in PR-AUC, reflecting their in-domain training advantage. The P-CHR AUC gap is considerably narrower (0.39--0.45 vs.\ 0.35--0.38), confirming that the domain advantage is smaller in actual deployment than offline evaluation suggests.

More fundamentally, the threshold-utility gap already exists at the retriever level. All 9 retrievers exhibit operational gaps, with $\Delta_{\text{op}}$ ranging from 0.26 to 0.39, yielding recoverable threshold-utility gaps between 0.07 and 0.20. Despite lower absolute P-CHR AUC, general-purpose retrievers retain a comparable fraction of offline quality, suggesting the inefficiency is similar across model families. Retriever scores, despite being the output of well-trained embedding models, are not well-placed for the binary threshold decision that a cache requires.

\begin{table*}[h]
\centering
\small
\begin{tabular}{lrrrrr}
\toprule
Retriever & PR-AUC & P-CHR AUC & $\Delta_{\text{op}}$ & $\Delta_{\text{util}}$ & ORR \\
\midrule
\multicolumn{6}{c}{\textit{Domain-specific}} \\
LangCache-Embed-v3   & \textbf{0.833} & \textbf{0.445} & 0.388 & 0.195 & 0.534 \\
LangCache-Embed-v2~\cite{gill2025advancingsemanticcachingllms}             & 0.754 & 0.402 & 0.352 & 0.159 & 0.533 \\
LangCache-Embed-v1~\cite{gill2025advancingsemanticcachingllms}             & 0.738 & 0.394 & 0.344 & 0.151 & 0.534 \\
\midrule
\multicolumn{6}{c}{\textit{General-purpose}} \\
BGE-base-en-v1.5~\cite{xiao2023cpack}               & 0.660 & 0.372 & 0.288 & 0.095 & 0.564 \\
GTE-ModernBERT-base~\cite{zhang-etal-2024-mgte}            & 0.649 & 0.380 & 0.269 & 0.076 & 0.586 \\
Jina-Embeddings-v2-base-en~\cite{gunther2023jina}     & 0.646 & 0.372 & 0.274 & 0.081 & 0.575 \\
Nomic-embed-text-v1.5~\cite{nussbaum2025nomic}          & 0.633 & 0.373 & 0.260 & 0.067 & 0.590 \\
E5-base-v2~\cite{wang2022e5}                     & 0.632 & 0.358 & 0.273 & 0.081 & 0.567 \\
Snowflake-Arctic-Embed-m-v2.0~\cite{yu2025arcticembed}  & 0.620 & 0.354 & 0.266 & 0.073 & 0.571 \\
\bottomrule
\end{tabular}
\caption{\textbf{Retriever baselines at $K=50$.} $\Delta_{\text{op}} = \text{PR-AUC} - \text{P-CHR AUC}$; $\Delta_{\text{util}} = \max(0, \Delta_{\text{op}} - \Delta_{\text{str}})$ estimates the recoverable threshold-utility component; ORR $=$ P-CHR AUC / PR-AUC estimates the operational retention. Every retriever carries a positive threshold-utility gap, indicating that retriever scores themselves are mis-placed for threshold-based cache decisions. Highest PR-AUC and P-CHR AUC in \textbf{bold}.}
\label{tab:retriever-baselines}
\end{table*}

\subsection{The Operational Gap}
\label{sec:results-calibration}

Table~\ref{tab:calibration-gap} reports PR-AUC, P-CHR AUC, $\Delta_{\text{op}}$, $\Delta_{\text{util}}$, and ORR for all 10 rerankers. We group models by behavioral cluster and average metrics across all 9 retrievers. The average retriever P-CHR AUC baseline is 0.383. We show that reranking rarely improves deployment quality: only ColBERTv2.0 (0.403) and GTE-Reranker (0.390) clear this average baseline, and against our strongest retriever (LangCache-Embed-v3, 0.445) no reranker improves at all. Models with the highest PR-AUC are not the best deployment choices; they are often the worst. Scaling the training data 38$\times$ from the v1 to the v2 reranker generation does not improve P-CHR AUC for either objective (MNRL: $0.353 \to 0.330$; BCE: $0.290 \to 0.171$), so the threshold-utility gap is set by the training objective, not data scale. Appendix~\ref{sec:curves} contrasts the PR and P-CHR curves for all rerankers, visualizing how the offline ranking reshuffles at deployment.

\paragraph{BCE training creates a model selection trap.} Both BCE-trained LangCache rerankers achieve high PR-AUC (0.816 and 0.748) yet P-CHR AUC of only 0.290 and 0.171, retaining 36\% and 23\% of their offline quality in deployment (ORR 0.356 and 0.229). Their threshold-utility gaps are 0.33 and 0.38 respectively, well above the structural gap, confirming that the score compression from BCE training leaves scores badly placed for the threshold decision. A practitioner selecting rerankers by PR-AUC would rank these models highly; P-CHR AUC inverts the ranking and reveals they are the worst-performing.

\paragraph{MNRL reduces but does not eliminate the threshold-utility gap.} MNRL rerankers improve on BCE substantially: P-CHR AUC of 0.353 and 0.330 compared to BCE's 0.290 and 0.171, retaining 41--43\% of offline quality (ORR 0.429 and 0.410), but both remain below the retriever baseline. Threshold-utility gaps of 0.28 are smaller than BCE's, but still remain significant. MNRL training learns relative ordering rather than absolute probabilities, which produces better-placed scores without fully solving the threshold-utility problem.

\paragraph{The gap extends beyond our models.} General-purpose cross-encoders exhibit similar behavior: GTE-Reranker-ModernBERT-base achieves PR-AUC of 0.712 yet P-CHR AUC of 0.390, only just clearing the average retriever baseline despite imposing reranking computation. ms-marco-MiniLM-L12-v2 sits mid-pack (P-CHR AUC 0.346, $\Delta_{\text{util}}$ 0.027), a modest gap but still below the retriever baseline. The threshold-utility gap is a property of how cross-encoder training interacts with threshold-based decisions, not an artifact of any particular model family or training dataset.

\paragraph{ColBERT-family models are the exception.} All four models achieve the lowest operational gaps (0.11--0.16) and the highest ORR values (0.685--0.782), despite the lowest PR-AUC (0.515--0.520). Their P-CHR AUC values cluster near or above the retriever baseline, and their $\Delta_{\text{util}} = 0$ indicates the gap is entirely structural, so no further recovery is possible. The inversion is sharpest for ColBERTv2.0, discussed in \S\ref{sec:results-colbert}.

\begin{table*}[t]
\centering
\small
\begin{tabular}{lrrrrr}
\toprule
Reranker & PR-AUC & P-CHR AUC & $\Delta_{\text{op}}$ & $\Delta_{\text{util}}$ & ORR \\
\midrule
\multicolumn{6}{c}{\textit{ColBERT-family}} \\
ColBERTv2.0~\cite{santhanam-etal-2022-colbertv2}                    & 0.515 & \textbf{0.403} & 0.112 & 0     & 0.782 \\
Reason-ModernColBERT~\cite{chaffin2025reasonmoderncolbert}           & 0.520 & 0.382          & 0.138 & 0     & 0.734 \\
ColBERT-Zero~\cite{chaffin2026colbertzero}                   & 0.518 & 0.375          & 0.143 & 0     & 0.724 \\
GTE-ModernColBERT-v1~\cite{chaffin2025gtemoderncolbert}           & 0.517 & 0.354          & 0.163 & 0     & 0.685 \\
\midrule
\multicolumn{6}{c}{\textit{General cross-encoders}} \\
GTE-Reranker-ModernBERT-base~\cite{zhang-etal-2024-mgte}   & 0.712 & 0.390 & 0.322 & 0.129 & 0.548 \\
ms-marco-MiniLM-L12-v2~\cite{wang2020minilm}         & 0.565 & 0.346 & 0.219 & 0.027 & 0.612 \\
\midrule
\multicolumn{6}{c}{\textit{LangCache Rerankers (MNRL)}} \\
LangCache-Reranker-v1-MNRL     & \textbf{0.824} & 0.353 & 0.471 & 0.278 & 0.429 \\
LangCache-Reranker-v2-MNRL     & 0.804 & 0.330 & 0.474 & 0.281 & 0.410 \\
\midrule
\multicolumn{6}{c}{\textit{LangCache Rerankers (BCE)}} \\
LangCache-Reranker-v1-BCE      & 0.816 & 0.290 & 0.526 & 0.333 & 0.356 \\
LangCache-Reranker-v2-BCE      & 0.748 & 0.171 & 0.577 & 0.384 & 0.229 \\
\bottomrule
\end{tabular}
\caption{\textbf{The operational gap across all 10 rerankers.} Metrics averaged across 9 retrievers. P-CHR AUC is the reranker deployment metric. $\Delta_{\text{util}} = \max(0, \Delta_{\text{op}} - \Delta_{\text{str}})$; ColBERT-family models have $\Delta_{\text{util}} = 0$ because their $\Delta_{\text{op}} \leq \Delta_{\text{str}}$, indicating the gap is entirely structural. ORR $=$ P-CHR AUC / PR-AUC; theoretical ceiling is $p(1-\ln p) \approx 0.809$. ColBERTv2.0 and GTE-Reranker improve P-CHR AUC over the average retriever baseline (0.383). Full per-retriever results are in Appendix~\ref{sec:full-results}. Highest PR-AUC and P-CHR AUC in \textbf{bold}.}
\label{tab:calibration-gap}
\end{table*}

\subsection{The ColBERT Inversion}
\label{sec:results-colbert}

ColBERTv2.0 presents the sharpest illustration that PR-AUC fails as a model selection criterion for deployment. It achieves the lowest PR-AUC of any reranker (0.515, barely above the positive rate $p = 0.45$) yet the highest P-CHR AUC (0.403) and is one of only two rerankers (with GTE-Reranker) to clear the average retriever baseline. Its operational gap of 0.112 is the smallest of all rerankers, compared to 0.471--0.577 for the high-PR-AUC MNRL and BCE models. A practitioner selecting by PR-AUC would rank ColBERTv2.0 last; ranked by P-CHR AUC, it is first. Its ORR of 0.782 is the highest of all rerankers, approaching the structural ceiling of 0.809, and reflects near-optimal score placement for threshold-based decisions. Yet even ColBERTv2.0's P-CHR AUC of 0.403 falls short of the best retriever alone: LangCache-Embed-v3 achieves 0.445 without any reranking.

This inversion is explained by ColBERT's scoring mechanism. Rather than producing a single logit, ColBERT aggregates token-level MaxSim scores and softmax-normalizes them across the $K$ retrieved candidates, yielding $\hat{s}(q) \in [0,1]$ with natural spread relative to the threshold. The relative score placement within the pool is sufficient for the binary cache decision, without requiring the model to estimate absolute probabilities. That the advantage is the pool-softmax normalization, not the token-level representation, is confirmed by ablation: stripping the softmax and scoring raw MaxSim with a per-pair sigmoid collapses ColBERTv2.0 from 0.403 to 0.020 P-CHR AUC, below every other reranker (Appendix~\ref{sec:normalization-ablation}). The same appendix shows the converse (applying the pool-softmax to the cross-encoders recovers much of their gap), so this normalization doubles as a training-free deployment fix. The remaining ColBERT-family models share this mechanism and similarly achieve low operational gaps; ColBERTv2.0 stands out within the family due to its stronger token-level representations, though at the strongest retriever GTE-Reranker overtakes it (Appendix~\ref{sec:full-results}). Appendix~\ref{sec:distribution-plots} makes this score placement visible across all rerankers, contrasting ColBERT's spread with the boundary collapse of the BCE models.

\paragraph{PR-AUC penalizes ColBERT unfairly.} PR-AUC ranks ColBERTv2.0 last, despite it being the best-deploying reranker on average. The penalty comes from the criterion, not the model: PR-AUC and P-CHR AUC read the same softmax-normalized MaxSim score but reward different properties of it. PR-AUC compares scores across all queries on a single scale, but ColBERT's score makes each score relative to its own candidate pool, so scores from different queries are not comparable. P-CHR AUC instead asks only whether the best candidate within a pool stands out enough to fire, which is what the softmax captures. The result is striking: a model whose pool-relative normalization makes it among the strongest choices for threshold-based caching is ranked last by standard offline metrics.

\section{Conclusion}
\label{sec:conclusion}

We recast semantic cache model selection as a threshold-utility problem rather than a ranking one: PR-AUC rewards ranking quality the deployment threshold never uses, while P-CHR AUC exposes the score placement it depends on. We introduce cache-aware metrics and decompose the operational gap they reveal into an irreducible structural component, fixed by the dataset positive rate, and a recoverable threshold-utility component. This gap is set mainly by the training objective, and measuring it is the first step to closing it. We release our datasets and models to support cache-aware evaluation, with practitioner recommendations in Appendix~\ref{sec:extended-discussion}.

\section*{Limitations}

Our evaluation is limited to English sentence-pair matching with a fixed candidate pool of $K=50$, and does not address multilingual caches or larger pools. Several further limitations exist: First, our cache-aware metrics require ground-truth relevance labels, which are often unavailable in production; monitoring threshold utility without labels remains an open problem. Second, both the structural gap and the threshold that maximizes deployment precision depend on the dataset positive rate, fixed at 45\% in our test set, so absolute P-CHR AUC values do not transfer across deployments with a different duplicate rate, even though the ranking of training objectives does. Third, we evaluate on a static test set, whereas production query distributions drift and require ongoing threshold maintenance. Finally, our cross-model comparisons span different score semantics. Our exact P-CHR AUC is invariant to any monotone rescaling of scores and so is commensurable across these semantics by construction; PR-AUC is not, and as we note for ColBERT (\S\ref{sec:results-colbert}), it can misrepresent a model whose scores are not produced as independent probabilities.

\section*{Ethical Considerations}

We adhere to the ACL Code of Ethics. Our experiments use only publicly available benchmark datasets; we collect no new user data and involve no human subjects, and we comply with all dataset licenses, releasing only license-compliant artifacts. Biases present in the source datasets may propagate through the trained retrieval and reranking models and affect cache-hit quality across demographic groups. We recommend evaluating with cache-aware metrics and validating the operating threshold on domain-representative data before user-facing deployment.

\section*{Acknowledgements}

We thank Redis for the compute resources and infrastructure support that enabled the large-scale training and evaluation in this work, and for the engineering and product feedback that grounded our metrics in the realities of production semantic caching. We are especially grateful to Warris Gill, whose prior work on the LangCache Embed models and the LLM Paraphrases dataset directly underpins this research, and to our colleagues on the Redis Vector Library and LangCache teams, both for the caching infrastructure this work is built on and the many discussions that shaped the deployment focus of this study. We also thank the authors and maintainers of the publicly available datasets that form the LangCache SentencePairs collection for their contributions to the research community and for making this evaluation possible. Finally, we thank the contributors to the open-source libraries whose tools made our training and evaluation pipeline practical.

\bibliography{custom}

\clearpage
\appendix

\section{Related Work}
\label{sec:related}

\paragraph{Semantic caching.}
Semantic caching reduces LLM inference costs by returning cached responses for semantically similar queries. Early work introduced embedding-based similarity matching for this purpose~\cite{bang-2023-gptcache, regmi2024gptsemanticcachereducing}, while later systems optimized deployment using domain-specific fine-tuned embeddings and vector databases~\cite{gill2025meancache, gill2025advancingsemanticcachingllms, wang2025categoryawaresemanticcachingheterogeneous,
yan2025contextcachecontextawaresemanticcache, liu2026semanticcachinglowcostllm}. Prior semantic caching work evaluates models using threshold-independent classification metrics such as PR-AUC, F1, or fixed-threshold precision~\cite{gill2025meancache, gill2025advancingsemanticcachingllms}. Recent work has recognized that static similarity thresholds fail to provide correctness guarantees, proposing per-prompt adaptive thresholds as a remedy~\cite{schroeder2026vcache, singh2026asynchronousverifiedsemanticcaching, biton2026exacthitscloseenough}.

\paragraph{Dense retrieval and two-stage pipelines.}
Bi-encoder models~\cite{reimers2019sentencebertsentenceembeddingsusing, karpukhin2020densepassageretrievalopendomain} independently encode queries and candidates into fixed-length vectors, enabling efficient approximate nearest neighbor search~\cite{johnson2017billionscalesimilaritysearchgpus, malkov2018efficientrobustapproximatenearest} at scale. Cross-encoder rerankers~\cite{nogueira2020passagererankingbert} jointly encode query--candidate pairs, allowing token-level attention to produce more accurate relevance scores at the cost of additional inference. Multi-vector models such as ColBERT~\cite{khattab2020colbertefficienteffectivepassage, santhanam-etal-2022-colbertv2} retain token-level representations while enabling precomputed index structures. Two-stage pipelines have established themselves as the current standard in open-domain QA and document retrieval.

\paragraph{Score calibration.}
Calibration measures the alignment between a model's predicted confidence and its actual correctness~\cite{guo2017calibrationmodernneuralnetworks}. Modern neural classifiers are often overconfident, and post-hoc methods such as temperature scaling~\cite{guo2017calibrationmodernneuralnetworks} and Platt scaling~\cite{platt1999probabilistic} are standard low-cost remedies applied after training. Temperature scaling applies a single learned scalar to the logit before sigmoid, while Platt scaling fits an affine transformation. Calibration has been studied extensively for classification tasks, but its role in threshold-based retrieval systems, where the deployment decision is a hard binary cut on the score, has received less attention.

\paragraph{Evaluation metrics for retrieval.}
Standard information retrieval metrics such as precision, recall, mean reciprocal rank, and nDCG measure ranking quality but are agnostic to absolute score magnitudes. ROC-AUC and PR-AUC are threshold-independent summaries that measure a model's ability to rank positives above negatives, but they are insensitive to whether those scores are concentrated in a deployable region of the score space. \citet{opitz-2024-schroedingers} show that AUC yields an optimistic notion of accuracy that can diverge substantially from threshold-based performance in text classification tasks.

\section{Dataset Curation}
\label{sec:dataset-curation}

This section provides comprehensive details on the construction, source composition, preprocessing, and statistics of the LangCache SentencePairs collection.

All versions share a common core evaluation set: v2 and v3 extend the v1 validation and test splits with the evaluation portions of their newly added sources, enabling direct comparison across training scales. Aggregate statistics are shown in Table~\ref{tab:dataset-stats}.

\begin{table}[t]
  \centering
  \small
  \begin{tabular}{lrrr}
    \toprule
    \textbf{Version} & \textbf{Train} & \textbf{Val} & \textbf{Test} \\
    \midrule
    v1 & 1,047,690 & 8,405 & 62,021 \\
    v2 & 8,184,872 & 8,405 & 72,021 \\
    v3 & 40,004,529 & 10,789 & 74,265 \\
    \bottomrule
  \end{tabular}
  \caption{\textbf{LangCache SentencePairs dataset statistics by version.}}
  \label{tab:dataset-stats}
\end{table}

\subsection{Source Datasets}

\begin{table*}[h]
  \centering
  \small
  \begin{tabular}{|p{3.8cm}|l|r|r|r|r|p{4cm}|}
    \hline
    \textbf{Source} & \textbf{Versions} & \textbf{Train} & \textbf{Val} & \textbf{Test} & \textbf{Total} & \textbf{Description} \\
    \hline
    \multicolumn{7}{|c|}{\textit{Core Benchmarks (v1, v2, v3)}} \\
    \hline
    PAWS~\cite{zhang2019pawsparaphraseadversariesword} & v1, v2, v3 & 645,652 & --- & 10,000 & 655,652 & Adversarial word-order changes, high lexical overlap \\
    QQP~\cite{iyer2017quora} & v1, v2, v3 & 363,846 & --- & 40,430 & 404,276 & Quora question duplicate detection \\
    PIT2015~\cite{xu-etal-2015-semeval} & v1, v2, v3 & 13,063 & 4,727 & 972 & 18,762 & Informal social media paraphrases \\
    PARADE~\cite{he-etal-2020-parade} & v1, v2, v3 & 7,550 & 1,275 & 1,357 & 10,182 & CS domain paraphrases requiring technical knowledge \\
    STS-B~\cite{cer-etal-2017-semeval} & v1, v2, v3 & 5,749 & 1,500 & 1,379 & 8,628 & Semantic similarity benchmark (binarized) \\
    SICK~\cite{marelli-etal-2014-sick} & v1, v2, v3 & 4,439 & 495 & 4,906 & 9,840 & Image caption pairs with compositional semantics \\
    APT~\cite{nighojkar-licato-2021-improving} & v1, v2, v3 & 3,723 & --- & 1,252 & 4,975 & Adversarial paraphrasing task \\
    MRPC~\cite{dolan-brockett-2005-automatically} & v1, v2, v3 & 3,668 & 408 & 1,725 & 5,801 & Microsoft Research news article paraphrases \\
    \hline
    \multicolumn{7}{|c|}{\textit{Synthetic Augmentation (v2 and v3)}} \\
    \hline
    LLM Paraphrases~\cite{gill2025advancingsemanticcachingllms} & v2, v3 & 7,065,497 & --- & 10,000 & 7,075,497 & LLM-generated diverse paraphrases across domains \\
    \hline
    \multicolumn{7}{|c|}{\textit{Large-Scale Expansion (v3 only)}} \\
    \hline
    ParaBank2~\cite{hu-etal-2019-large} & v3 & 19,425,235 & --- & --- & 19,425,235 & Large-scale neural MT paraphrases via back-translation \\
    ChatGPT Paraphrases~\cite{chatgpt_paraphrases_dataset} & v3 & 6,286,314 & --- & --- & 6,286,314 & ChatGPT-generated paraphrase variations \\
    ParaNMT-5M~\cite{wieting-gimpel-2018-paranmt} & v3 & 5,370,126 & --- & --- & 5,370,126 & Neural MT paraphrases (filtered subset) \\
    Opusparcus~\cite{creutz2018opensubtitlesparaphrasecorpus} & v3 & 1,500,000 & 1,455 & 1,445 & 1,502,900 & Movie subtitle paraphrases from translations \\
    TTIC31190~\cite{ttic-31190-paraphrase} & v3 & 1,000,000 & 818 & 805 & 1,001,623 & Kaggle paraphrase competition data \\
    Paraphrase Collections~\cite{Xu_2023} & v3 & 223,241 & --- & --- & 223,241 & Aggregated paraphrase datasets \\
    TaPaCo~\cite{scherrer-2020-tapaco} & v3 & 224,824 & --- & --- & 224,824 & Tatoeba paraphrase corpus \\
    Task275~\cite{wang2022supernaturalinstructionsgeneralizationdeclarativeinstructions} & v3 & 942 & 118 & 118 & 1,178 & Enhanced WSC paraphrase generation \\
    \hline
  \end{tabular}
  \caption{\textbf{Complete source overview across all LangCache SentencePairs versions}. MT = Machine Translation. Sources are grouped by version introduction: core benchmarks (v1), synthetic augmentation (v2), and large-scale expansion (v3). Val and Test columns show evaluation splits where available.}
  \label{tab:all-sources-comprehensive}
\end{table*}

Table~\ref{tab:all-sources-comprehensive} provides a complete overview of all sources across all versions. Sources are organized into three categories.

\paragraph{Core Benchmarks.} Eight foundational datasets introduced in v1 and retained in all later versions for consistent evaluation. Version 1 establishes a high-quality baseline that prioritizes linguistic variety, so that models learn robust representations generalizing beyond any single domain.

\paragraph{Synthetic Augmentation.} Large-scale LLM-generated paraphrases~\cite{gill2025advancingsemanticcachingllms} added in v2 and carried forward to v3, providing diverse phrasing across domains, achieving an 8$\times$ scale increase. The validation split remains unchanged from v1 (8,405 pairs). The test split expands to 72,021 pairs by adding 10,000 synthetic examples, ensuring evaluation covers both human-authored and LLM-generated paraphrases.

\paragraph{Large-Scale Expansion.} Eight additional sources totaling 34M pairs added to v3 only, representing a 38$\times$ scale increase over v1 and incorporating large-scale machine-translated paraphrases, movie subtitle paraphrases, and additional synthetic data. Version 3 expands validation to 10,789 pairs and test to 74,265 pairs, incorporating examples from newly added sources while preserving the core evaluation set from v1 for longitudinal comparison.

\subsection{Split Strategy}

Original train/validation/test splits from source datasets are preserved where available to prevent data leakage and maintain comparability with prior work. For sources without validation splits, all non-test data goes to training. All splits are mutually exclusive; no data leakage occurs between train, validation, and test.

\subsection{Preprocessing Pipeline}

All source datasets undergo consistent preprocessing before integration:

\begin{enumerate}
  \item \textbf{Schema Normalization.} Each source is converted to a uniform $(\texttt{sentence1}, \texttt{sentence2}, \texttt{label})$ format, and pairs with a missing sentence are dropped.

  \item \textbf{Label Binarization.} Graded and categorical labels are mapped to $\{0,1\}$ per source: STS-B similarity scores above $3.5$ (on a $0$--$5$ scale) are positive; SICK pairs are positive unless their entailment label is \textsc{contradiction}; PIT-2015 pairs are positive when at least four of five annotators judge them paraphrases; Opusparcus pairs are positive when their annotator score is at least $3.0$; and the remaining benchmarks (PAWS, QQP, MRPC, APT) use their native binary labels.

  \item \textbf{Deduplication.} After all sources are merged, exact duplicate $(\texttt{sentence1}, \texttt{sentence2})$ pairs are removed within each of the train, validation, and test splits. The version totals in Table~\ref{tab:dataset-stats} are therefore smaller than the sums of the per-source counts in Table~\ref{tab:all-sources-comprehensive}.
\end{enumerate}

\subsection{Quality Considerations and Limitations}

\paragraph{Label Noise.} Some sources include noisy labels, particularly: PAWS weakly-labeled pairs from large-scale automatic generation; machine-translated paraphrases (ParaBank2, ParaNMT) which may contain translation artifacts; and Opusparcus subtitle paraphrases which may reflect biased translation choices rather than true semantic equivalence.

\paragraph{Domain Imbalance.} In v3, ParaBank2 (19.4M pairs, 48.6\%) and synthetic data (13.4M pairs, 33.5\%) dominate the training set. Users seeking balanced representation across domains should apply sampling or weighting strategies during training.

\paragraph{Synthetic Data Characteristics.} LLM-generated and machine-translated paraphrases may exhibit different distributional properties than human-authored text, including more formal language, less idiomatic expression, and potential amplification of training data biases from source LLMs.

\paragraph{Granularity Loss.} STS-B's continuous similarity scores (0--5 range) are binarized to $\{0,1\}$ labels. The original dataset should be consulted for fine-grained similarity regression tasks.

\paragraph{English-Only.} All datasets contain exclusively English text. Multilingual or cross-lingual applications require separate data.

\paragraph{Class Imbalance.} The test set maintains a 45\% positive / 55\% negative split, which may not reflect production semantic cache deployments. Practitioners deploying in low-duplicate environments should evaluate threshold selection using domain-representative data.

\section{Training Configuration}
\label{sec:training-config}

\paragraph{LangCache-Embed-v3 Training.} We fine-tune all-MiniLM-L6-v2~\cite{wang2020minilm, reimers2019sentencebertsentenceembeddingsusing}, a 6-layer, 22.6M-parameter MiniLM encoder, with mean pooling and a 128-token context window producing 384-dimensional embeddings. We train on LangCache SentencePairs v2 (8M pairs) in BF16 using an in-batch contrastive objective: the ArcFace additive angular margin loss~\cite{deng2019arcface} with cosine similarity and scale $20$, which pulls each anchor toward its positive while pushing it away from its mined hard negative and all other in-batch examples. Embeddings are unit-normalized and compared by cosine similarity at inference. The optimization recipe follows Table~\ref{tab:reranker-training-config}, with the maximum sequence length reduced to $128$ tokens to match the encoder's context.

\paragraph{LangCache Reranker Training.} The LangCache Rerankers fine-tune GTE-Reranker-ModernBERT-base~\cite{zhang-etal-2024-mgte} under the hyperparameters in Table~\ref{tab:reranker-training-config}, identical across training objectives and across the v1 (1M pairs) and v3 (40M pairs) training scales, so that any difference in behavior reflects the objective or scale rather than the optimization recipe. 

We choose systematically different objectives for our comparison: \textbf{Binary Cross-Entropy (BCE)} treats reranking as binary classification, minimizing cross-entropy loss against hard relevance labels. The model directly outputs probabilities $\hat{p} = \sigma(z)$ where $z$ is the final logit. The v3 expansion corpora contribute positive pairs only, leaving the merged training set positive-skewed. BCE therefore uses a class-balanced cross-entropy that weights the positive class by $N_{\text{neg}}/N_{\text{pos}}$, the negative-to-positive ratio of training pairs. This equalizes the total loss contribution of the two classes, so the BCE score-placement behavior we report reflects the objective rather than label imbalance. \textbf{Multiple Negatives Ranking Loss (MNRL)}~\cite{reimers2019sentencebertsentenceembeddingsusing} is a contrastive objective over triplets (anchor, positive, negative), with hard negatives sampled from labeled negative pairs per anchor and additional in-batch negatives. MNRL learns relative ordering to rank positives higher than negatives, rather than estimating absolute probability boundaries.

Checkpoints for both retriever and reranker are selected by validation F1-score. Training used AWS \texttt{g5.12xlarge} instances (4$\times$ NVIDIA A10G) with PyTorch Distributed Data Parallel (DDP).

\paragraph{Validation monitoring.} Training ran two evaluators every $1{,}000$ training steps. A classification evaluator computed precision, recall, F1, and accuracy on the validation split; F1 drives checkpoint selection. A cache utilization evaluator computed PR-AUC, P-CHR AUC, and distributional overlap on the same split, providing early visibility into score placement without influencing the checkpoint decision.

\begin{table}[t]
  \centering
  \small
  \begin{tabular}{ll}
    \toprule
    \textbf{Parameter} & \textbf{Value} \\
    \midrule
    Batch size & 48 \\
    Learning rate & $2 \times 10^{-4}$ \\
    Max sequence length & 512 tokens \\
    Gradient clipping & 1.0 \\
    Scheduler & Linear decay \\
    Warmup ratio & 0.1 \\
    Weight decay & 0.003 \\
    Epochs & 5 \\
    Optimizer & AdamW \\
    Selection metric & Validation F1 \\
    \bottomrule
  \end{tabular}
  \caption{\textbf{Training hyperparameters for the LangCache Rerankers}, shared across BCE/MNRL objectives and both training scales.}
  \label{tab:reranker-training-config}
\end{table}

\paragraph{Why we do not train ColBERT-family rerankers.} We evaluate ColBERT-family models off the shelf but do not fine-tune our own multi-vector reranker. Multi-vector scoring retains per-token representations for every candidate, and the resulting index and late-interaction overhead often preclude its use in latency- and memory-constrained production semantic caches. We therefore restrict our trained models to single-vector cross-encoders and treat ColBERT as an informative external point of comparison.

\section{Metric Definitions and Derivations}
\label{sec:metric-derivations}

\subsection{Classification Metrics}

For threshold $\tau$, standard classification outcomes are:
\begin{equation}
\begin{aligned}
\text{TP} &= |\{q : y_q{=}1 \wedge s(q,c^*){\geq}\tau\}| \\
\text{FP} &= |\{q : y_q{=}0 \wedge s(q,c^*){\geq}\tau\}| \\
\text{FN} &= |\{q : y_q{=}1 \wedge s(q,c^*){<}\tau\}|  \\
\text{TN} &= |\{q : y_q{=}0 \wedge s(q,c^*){<}\tau\}|
\end{aligned}
\end{equation}

\noindent\textbf{Precision}$(\tau) = \text{TP}/(\text{TP}+\text{FP})$\quad measures the fraction of cache responses that are correct.
\textbf{Recall}$(\tau) = \text{TP}/(\text{TP}+\text{FN})$\quad measures the fraction of positive queries served from cache.

\subsection{Why Cache Hit Ratio, Not Recall}
\label{sec:chr-vs-recall}

In a search scenario, computing the recall of a system is non-trivial. Let $Q$ denote the set of incoming queries and $C$ the set of cached entries. Relevance is defined by a sparse binary matrix $R \in \{0,1\}^{|Q| \times |C|}$, where $R_{qc}=1$ iff cache entry $c$ is semantically valid for query $q$. A semantic cache returns a hit for query $q$ if there exists at least one cached entry whose score exceeds the threshold $\tau$.

Recall, however, requires knowing whether the system retrieved a relevant entry among all relevant cached entries. For a cache miss, observing that no returned candidate exceeds the threshold does not reveal whether the miss is a false negative or a true negative: distinguishing these cases requires determining whether $\exists c \in C$ such that $R_{qc}=1$, which in the worst case requires labeling the entire row $R_{q,:}$. Because $|C|$ can be large and the relevance matrix is extremely sparse, this exhaustive labeling is prohibitively expensive and makes recall difficult to estimate reliably in production.

Cache Hit Ratio avoids this problem by measuring the observable event that the cache returns any candidate:
\begin{equation}
  \text{CHR}(\tau) = \frac{1}{|Q|} \sum_{q \in Q} \mathbb{I}\left[\exists c \in C : s(q,c) \geq \tau\right]
\end{equation}
Unlike recall, CHR does not require knowledge of unobserved relevant entries for missed queries. It is therefore directly measurable from system logs and captures the operational quantity that matters for semantic caching: the fraction of LLM calls avoided. When paired with precision over accepted hits, P-CHR curves measure the production tradeoff between correctness and cache utilization without requiring complete relevance annotations over the full query--cache matrix.

\subsection{Deployment Precision as VCHR/CHR}

\textbf{Valid Cache Hit Ratio (VCHR).}
\begin{equation}
\begin{aligned}
  \text{VCHR}(\tau) = \frac{1}{N}\,\big|\{q : \;&\hat{s}(q) \geq \tau \;\wedge\; \hat{c}(q) = c^* \\
  &\wedge\; y_q = 1\}\big|
\end{aligned}
\end{equation}
VCHR counts only valid fires. In this setting, the cache-decision precision $\text{Precision}(\tau) = \text{VCHR}(\tau) / \text{CHR}(\tau)$: VCHR is the product of cache utilization and correctness. This deployment precision is computed on the top-1 score $\hat{s}(q)$ and is distinct from the classification precision of \S\ref{sec:metrics-classification}, which thresholds the ground-truth score $s(q, c^*)$ rather than the top-1 score $\hat{s}(q)$.

At threshold $\tau$, the cache fires on every query with $\hat{s}(q) \geq \tau$, so the number of fires is $N \cdot \text{CHR}(\tau)$. A fire is correct exactly when it is valid, i.e.\ when the top-ranked candidate is the ground truth and the query is a genuine positive ($\hat{c}(q)=c^*$ and $y_q=1$), of which there are $N \cdot \text{VCHR}(\tau)$. The deployment precision is the fraction of fires that are correct,
\begin{equation}
  \text{Precision}(\tau) = \frac{N\cdot\text{VCHR}(\tau)}{N\cdot\text{CHR}(\tau)} = \frac{\text{VCHR}(\tau)}{\text{CHR}(\tau)}
\end{equation}
\noindent which is integrated against CHR to form P-CHR AUC and against VCHR to form P-VCHR AUC. 

\textbf{Precision-VCHR curves} substitute VCHR on the x-axis, giving a more conservative measure that accounts only for valid fires:
\begin{equation}
  \text{P-VCHR AUC} = \int_{0}^{p} \text{Precision}\bigl(\text{VCHR}^{-1}(v)\bigr)\, dv
\end{equation}

\subsection{Structural Gap Derivation}

Consider a perfect ranker, one that assigns every positive query a higher top-1 score than every negative query, and whose top-ranked candidate for each positive is its ground truth. Parametrize the P-CHR curve by $c = \text{CHR} \in [0,1]$, the fraction of queries that fire. As $\tau$ decreases, queries fire in score order. While $c \leq p$, every firing query is one of the $Np$ positives and is served correctly, so $\text{VCHR}=c$ and
\begin{equation}
  \text{Precision}(c) = \frac{\text{VCHR}}{\text{CHR}} = \frac{c}{c} = 1
\end{equation}
Once $c > p$, all $Np$ positives have already fired ($\text{VCHR}=p$) and every additional fire is a negative, so
\begin{equation}
  \text{Precision}(c) = \frac{p}{c}
\end{equation}
The best achievable P-CHR AUC is therefore
\begin{equation}
\begin{aligned}
  \int_0^1 \text{Precision}(c)\, dc &= \int_0^p 1\, dc + \int_p^1 \frac{p}{c}\, dc \\
  &= p + p\bigl(\ln 1 - \ln p\bigr) \\
  &= p\,(1 - \ln p)
\end{aligned}
\end{equation}
Since PR-AUC is at most $1$, even a perfect ranker incurs the structural gap
\begin{equation}
  \Delta_{\text{str}} = 1 - p\,(1 - \ln p)
\end{equation}
which evaluates to $\Delta_{\text{str}} \approx 0.191$ at $p = 0.45$ and depends only on the positive rate, not on the model. The derivation assumes perfect retrieval recall; retrieval misses lower the achievable P-CHR AUC further but do not change $\Delta_{\text{str}}$.

\subsection{Properties of ORR}

The same argument bounds P-CHR AUC by $p(1-\ln p) \approx 0.809$ for \emph{any} model, since the perfect ranker maximizes precision at every $c$. Hence a perfectly-placed model with PR-AUC $=1$ attains $\text{ORR} = 0.809$, the structural ceiling. ORR is most informative for models whose PR-AUC lies well above the random baseline $p$. A high P-CHR AUC requires firing positives first with high precision, which in turn requires strong ranking, so P-CHR AUC and PR-AUC move together and ORR cleanly measures how efficiently ranking quality is converted into deployment precision. As PR-AUC approaches $p$, the structural bound becomes loose and ORR is no longer monotone in model quality. This low-PR-AUC regime is exactly where $\Delta_{\text{op}} \leq \Delta_{\text{str}}$ and the clamp $\Delta_{\text{util}} = \max(0, \Delta_{\text{op}} - \Delta_{\text{str}})$ sets the threshold-utility gap to zero, as it does for the ColBERT-family models.

\section{Full Results}
\label{sec:full-results}

Tables~\ref{tab:full-prauc}, \ref{tab:full-pchr}, and \ref{tab:full-pvchr} report PR-AUC, P-CHR AUC, and P-VCHR AUC for all 90 retriever--reranker combinations at $K=50$. The main text reports per-reranker averages across retrievers. The behavioral clustering is stable across all nine retrievers: ColBERT-family models hold the highest or near-highest P-CHR AUC despite the lowest PR-AUC (GTE-Reranker edges ahead only under the strongest retriever), while BCE rerankers hold the lowest P-CHR AUC despite high PR-AUC, and MNRL sits in between. P-VCHR AUC, which counts only valid fires and is bounded by $p$, preserves the same ordering.

\begin{table*}[t]
\centering
\small
\setlength{\tabcolsep}{4pt}
\begin{tabular}{lrrrrrrrrrr}
\toprule
 & \multicolumn{4}{c}{\textit{ColBERT-family}} & \multicolumn{2}{c}{\textit{Cross-encoders}} & \multicolumn{2}{c}{\textit{LangCache MNRL}} & \multicolumn{2}{c}{\textit{LangCache BCE}} \\
\cmidrule(lr){2-5}\cmidrule(lr){6-7}\cmidrule(lr){8-9}\cmidrule(lr){10-11}
Retriever & CBv2 & Reason-CB & CB-0 & GTE-CB & GTE-Rank & MiniLM & v1-MNRL & v2-MNRL & v1-BCE & v2-BCE \\
\midrule
\multicolumn{11}{c}{\textit{Domain-specific}} \\
LC Embed v3          &0.523 & 0.525 & 0.525 & 0.522 & 0.734 & 0.592 & \textbf{0.867} & 0.838 & 0.852 & 0.790 \\
LC Embed v2 &0.518 & 0.518 & 0.516 & 0.516 & 0.719 & 0.571 & \textbf{0.836} & 0.818 & 0.825 & 0.748 \\
LC Embed v1 &0.520 & 0.519 & 0.517 & 0.517 & 0.718 & 0.569 & \textbf{0.834} & 0.815 & 0.825 & 0.745 \\
\midrule
\multicolumn{11}{c}{\textit{General-purpose}} \\
BGE & 0.512 & 0.520 & 0.518 & 0.517 & 0.709 & 0.562 & \textbf{0.813} & 0.795 & 0.808 & 0.734 \\
GTE-MB & 0.514 & 0.521 & 0.516 & 0.517 & 0.709 & 0.560 & \textbf{0.816} & 0.797 & 0.812 & 0.739 \\
Jina & 0.512 & 0.519 & 0.516 & 0.516 & 0.709 & 0.559 & \textbf{0.817} & 0.795 & 0.808 & 0.733 \\
Nomic & 0.513 & 0.520 & 0.518 & 0.517 & 0.708 & 0.558 & \textbf{0.816} & 0.795 & 0.808 & 0.751 \\
E5 & 0.513 & 0.519 & 0.519 & 0.516 & 0.705 & 0.560 & \textbf{0.810} & 0.792 & 0.805 & 0.748 \\
Arctic & 0.513 & 0.520 & 0.517 & 0.517 & 0.701 & 0.555 & \textbf{0.807} & 0.788 & 0.802 & 0.741 \\
\bottomrule
\end{tabular}
\caption{\textbf{PR-AUC for all 90 combinations.} Rerankers: CBv2 = ColBERTv2.0, Reason-CB = Reason-ModernColBERT, CB-0 = ColBERT-Zero, GTE-CB = GTE-ModernColBERT-v1, GTE-Rank = GTE-Reranker-ModernBERT-base, MiniLM = ms-marco-MiniLM-L12-v2, v1-MNRL/v2-MNRL = LangCache MNRL, v1-BCE/v2-BCE = LangCache BCE. Retrievers: LC Embed v3/v2/v1 = LangCache-Embed v3/v2/v1, GTE-MB = GTE-ModernBERT-base, Arctic = Snowflake-Arctic-Embed-m-v2.0. Highest value per retriever (row) in \textbf{bold}. The MNRL rerankers achieve the highest PR-AUC for every retriever, yet rank poorly on deployment (Table~\ref{tab:full-pchr}), illustrating the operational gap.}
\label{tab:full-prauc}
\end{table*}

\begin{table*}[t]
\centering
\small
\setlength{\tabcolsep}{4pt}
\begin{tabular}{lrrrrrrrrrr}
\toprule
 & \multicolumn{4}{c}{\textit{ColBERT-family}} & \multicolumn{2}{c}{\textit{Cross-encoders}} & \multicolumn{2}{c}{\textit{LangCache MNRL}} & \multicolumn{2}{c}{\textit{LangCache BCE}} \\
\cmidrule(lr){2-5}\cmidrule(lr){6-7}\cmidrule(lr){8-9}\cmidrule(lr){10-11}
Retriever & CBv2 & Reason-CB & CB-0 & GTE-CB & GTE-Rank & MiniLM & v1-MNRL & v2-MNRL & v1-BCE & v2-BCE \\
\midrule
\multicolumn{11}{c}{\textit{Domain-specific}} \\
LC Embed v3          & 0.405 & 0.384 & 0.379 & 0.357 & \textbf{0.422} & 0.383 & 0.392 & 0.363 & 0.337 & 0.232 \\
LC Embed v2 & \textbf{0.398} & 0.378 & 0.372 & 0.352 & 0.396 & 0.352 & 0.362 & 0.337 & 0.296 & 0.142 \\
LC Embed v1 & \textbf{0.399} & 0.379 & 0.373 & 0.352 & 0.393 & 0.350 & 0.360 & 0.334 & 0.298 & 0.141 \\
\midrule
\multicolumn{11}{c}{\textit{General-purpose}} \\
BGE & \textbf{0.403} & 0.383 & 0.376 & 0.355 & 0.382 & 0.335 & 0.341 & 0.321 & 0.277 & 0.149 \\
GTE-MB & \textbf{0.404} & 0.383 & 0.374 & 0.355 & 0.387 & 0.341 & 0.347 & 0.329 & 0.288 & 0.164 \\
Jina & \textbf{0.404} & 0.382 & 0.374 & 0.354 & 0.385 & 0.339 & 0.346 & 0.323 & 0.279 & 0.151 \\
Nomic & \textbf{0.405} & 0.383 & 0.375 & 0.355 & 0.390 & 0.345 & 0.351 & 0.325 & 0.285 & 0.189 \\
E5 & \textbf{0.404} & 0.382 & 0.376 & 0.355 & 0.377 & 0.333 & 0.339 & 0.321 & 0.274 & 0.186 \\
Arctic & \textbf{0.404} & 0.384 & 0.376 & 0.355 & 0.381 & 0.335 & 0.342 & 0.315 & 0.278 & 0.183 \\
\bottomrule
\end{tabular}
\caption{\textbf{P-CHR AUC for all 90 combinations.} Column and row abbreviations as in Table~\ref{tab:full-prauc}; highest value per retriever (row) in \textbf{bold}. ColBERT-family P-CHR AUC is highest for eight of nine retrievers (GTE-Reranker edges ahead under LangCache-Embed-v3) and the most stable across retrievers; BCE is lowest.}
\label{tab:full-pchr}
\end{table*}

\begin{table*}[t]
\centering
\small
\setlength{\tabcolsep}{4pt}
\begin{tabular}{lrrrrrrrrrr}
\toprule
 & \multicolumn{4}{c}{\textit{ColBERT-family}} & \multicolumn{2}{c}{\textit{Cross-encoders}} & \multicolumn{2}{c}{\textit{LangCache MNRL}} & \multicolumn{2}{c}{\textit{LangCache BCE}} \\
\cmidrule(lr){2-5}\cmidrule(lr){6-7}\cmidrule(lr){8-9}\cmidrule(lr){10-11}
Retriever & CBv2 & Reason-CB & CB-0 & GTE-CB & GTE-Rank & MiniLM & v1-MNRL & v2-MNRL & v1-BCE & v2-BCE \\
\midrule
\multicolumn{11}{c}{\textit{Domain-specific}} \\
LC Embed v3          & 0.124 & 0.114 & 0.112 & 0.102 & \textbf{0.132} & 0.112 & 0.116 & 0.100 & 0.085 & 0.048 \\
LC Embed v2 & \textbf{0.121} & 0.111 & 0.109 & 0.100 & 0.118 & 0.098 & 0.100 & 0.087 & 0.068 & 0.021 \\
LC Embed v1 & \textbf{0.121} & 0.111 & 0.109 & 0.100 & 0.116 & 0.097 & 0.099 & 0.086 & 0.068 & 0.021 \\
\midrule
\multicolumn{11}{c}{\textit{General-purpose}} \\
BGE & \textbf{0.123} & 0.113 & 0.111 & 0.101 & 0.111 & 0.091 & 0.090 & 0.080 & 0.060 & 0.022 \\
GTE-MB & \textbf{0.123} & 0.113 & 0.110 & 0.101 & 0.114 & 0.094 & 0.093 & 0.083 & 0.064 & 0.026 \\
Jina & \textbf{0.123} & 0.112 & 0.110 & 0.100 & 0.112 & 0.093 & 0.093 & 0.080 & 0.061 & 0.022 \\
Nomic & \textbf{0.124} & 0.113 & 0.110 & 0.101 & 0.114 & 0.095 & 0.095 & 0.081 & 0.063 & 0.032 \\
E5 & \textbf{0.123} & 0.112 & 0.110 & 0.101 & 0.108 & 0.090 & 0.089 & 0.079 & 0.059 & 0.032 \\
Arctic & \textbf{0.123} & 0.113 & 0.111 & 0.101 & 0.110 & 0.091 & 0.090 & 0.077 & 0.060 & 0.032 \\
\bottomrule
\end{tabular}
\caption{\textbf{P-VCHR AUC for all 90 combinations.} Column and row abbreviations as in Table~\ref{tab:full-prauc}; highest value per retriever (row) in \textbf{bold}. P-VCHR AUC counts only valid fires and is bounded by $p \approx 0.45$; it preserves the P-CHR AUC ordering.}
\label{tab:full-pvchr}
\end{table*}

\section{Post-Hoc Calibration}
\label{sec:calibration-fitting}

BCE training minimizes cross-entropy against binary relevance labels, giving model outputs a direct probabilistic interpretation: $\hat{p}(q, c) \approx P(\text{relevant} \mid q, c)$. Post-hoc recalibration is therefore principled for BCE models: a monotonic transformation of the score can correct systematic deviations from the true posterior without altering rank order. MNRL models do not share this interpretation; their logits encode relative preference, not calibrated probabilities, and post-hoc calibration is not applicable. We nonetheless apply two standard calibration methods, \textbf{Temperature scaling}~\cite{guo2017calibrationmodernneuralnetworks} and \textbf{Platt scaling}~\cite{platt1999probabilistic}, to all four LangCache rerankers to test whether recalibration recovers deployment quality.

\textbf{Temperature scaling} introduces a single scalar $T > 0$, replacing $\sigma(z)$ with $\sigma(z / T)$. A value $T > 1$ spreads the score distribution away from its current concentration; $T < 1$ compresses it. \textbf{Platt scaling} fits a two-parameter logistic regression on the raw logits: $\sigma(az + b)$, allowing both rescaling and shifting of the score distribution. In both cases, parameters are fit by minimizing negative log-likelihood on the LangCache SentencePairs v3 validation split.

Table~\ref{tab:calib-params} lists the fitted parameters. The BCE models require large temperatures ($T = 9.78$ and $20.09$), confirming severe score compression: their logits must be divided by a large factor to spread the sigmoid output across $[0,1]$. The MNRL models require $T < 1$, indicating their scores are already spread. LangCache-Reranker-v2-BCE's extreme Platt parameters ($a=5.08,\, b=12.12$) reflect the same severe compression of its logits.

\begin{table}[t]
\centering
\small
\begin{tabular}{lrrrc}
\toprule
Reranker & $T$ & Platt $a$ & Platt $b$ & $\Delta$P-CHR AUC \\
\midrule
v1-BCE  & 9.78  & 0.17 & 2.59  & $0.000$ \\
v2-BCE  & 20.09 & 5.08 & 12.12 & $0.000$ \\
v1-MNRL & 0.67  & 0.98 & 2.22  & $0.000$ \\
v2-MNRL & 0.24  & 3.84 & 0.82  & $0.000$ \\
\bottomrule
\end{tabular}
\caption{\textbf{Post-hoc calibration parameters and their effect}. The BCE models require large $T$ ($9.78$, $20.09$) to undo severe score compression, whereas the MNRL models require $T < 1$, confirming their scores are already well spread. $\Delta$P-CHR AUC is the change in exact P-CHR AUC (averaged over 9 retrievers) under temperature or Platt scaling: both are strictly monotone and P-CHR AUC is rank-invariant, so it is exactly $0.000$ for every reranker; PR-AUC is likewise unchanged.}
\label{tab:calib-params}
\end{table}

\subsection{Effect on P-CHR AUC}
\label{sec:results-calibration-gain}

Because temperature and Platt scaling are strictly monotone in the score, and P-CHR AUC depends only on the \emph{order} of scores (\S\ref{sec:metrics-cache}), neither can change it. The calibration gain
\begin{equation}
  \text{Gain}_{\text{cal}} = \text{P-CHR AUC}_{\text{post}} - \text{P-CHR AUC}_{\text{pre}}
\end{equation}
is exactly $0.000$ for all four rerankers (Table~\ref{tab:calib-params}, last column), as is the change in PR-AUC. Post-hoc calibration therefore cannot recover deployment quality, however severe the score compression: the fitted temperatures confirm the BCE models are badly compressed ($T = 9.78, 20.09$), yet a monotone rescaling moves every operating point together and leaves the precision--CHR curve fixed. Recovering the gap requires either re-normalizing scores over the candidate pool with a softmax (Appendix~\ref{sec:normalization-ablation}) or a training objective that places scores well to begin with. Post-hoc probability calibration and threshold utility are different targets: a monotone map can make scores better probabilities without making them any more usable at a fixed threshold.

\section{Sensitivity to Candidate Pool Size}
\label{sec:k-sensitivity}

Table~\ref{tab:k-sensitivity} reports P-CHR AUC as a function of the candidate pool size $K$ for all rerankers paired with LangCache-Embed-v3. The behavior splits sharply by scoring mechanism. Cross-encoders that score each candidate independently (GTE-Rank, MiniLM, and the LangCache MNRL and BCE models) peak at $K=2$ and decline as $K$ grows: larger pools increase the likelihood that a high-scoring false positive will fire at any threshold, eroding precision. The ColBERT-family models behave differently because they softmax-normalize over the pool and cannot produce a usable distribution at $K=1$ (P-CHR AUC $=0$); they require a pool of at least two candidates and peak later. Their fates then diverge: only ColBERTv2.0 keeps improving as the pool grows, plateauing near $K=50$, whereas Reason-CB, ColBERT-0, and GTE-CB peak by $K=5$--$10$ and decline thereafter like the independently-scored models. The retriever baseline saturates at $K=2$.

This dependence means the headline comparison is $K$-specific. The baseline here is LangCache-Embed-v3's own P-CHR AUC ($0.445$), well above the nine-retriever average of $0.383$ used in \S\ref{sec:results-calibration}. Against this strong retriever, almost no reranker improves deployment quality at any $K$: the single exception in the entire table is GTE-Reranker at $K=2$ ($0.447$ vs.\ the retriever's $0.445$), a margin within noise, and at $K=50$ every reranker (ColBERTv2.0 included) falls short of $0.445$. MNRL outperforms BCE at every $K$, so the central comparison of training objectives does not depend on the pool size. The ColBERT-family, in contrast, is unusable at $K = 1$ and becomes competitive only once the pool is large enough to normalize over.

\begin{table*}[t]
\centering
\small
\setlength{\tabcolsep}{6pt}
\begin{tabular}{lrrrrrrrr}
\toprule
Model & \multicolumn{8}{c}{$K$} \\
\cmidrule(lr){2-9}
 & 1 & 2 & 5 & 10 & 20 & 30 & 40 & 50 \\
\midrule
\multicolumn{9}{c}{\textit{Retriever baseline}} \\
LangCache-Embed-v3                & 0.381 & \textbf{0.445} & 0.445 & 0.445 & 0.445 & 0.445 & 0.445 & 0.445 \\
\midrule
\multicolumn{9}{c}{\textit{ColBERT-family}} \\
ColBERTv2.0                  & 0.000 & 0.371 & 0.387 & 0.396 & 0.401 & 0.403 & 0.404 & \textbf{0.405} \\
Reason-ModernColBERT         & 0.000 & 0.385 & 0.398 & \textbf{0.399} & 0.394 & 0.390 & 0.387 & 0.384 \\
ColBERT-Zero                 & 0.000 & 0.394 & \textbf{0.409} & 0.401 & 0.391 & 0.385 & 0.381 & 0.379 \\
GTE-ModernColBERT-v1         & 0.000 & 0.394 & \textbf{0.397} & 0.380 & 0.368 & 0.363 & 0.359 & 0.357 \\
\midrule
\multicolumn{9}{c}{\textit{General cross-encoders}} \\
GTE-Reranker-ModernBERT-base & 0.390 & \textbf{0.447} & 0.432 & 0.428 & 0.425 & 0.424 & 0.423 & 0.422 \\
ms-marco-MiniLM-L12-v2       & 0.361 & \textbf{0.420} & 0.400 & 0.395 & 0.390 & 0.387 & 0.385 & 0.383 \\
\midrule
\multicolumn{9}{c}{\textit{LangCache Rerankers (MNRL)}} \\
LangCache-Reranker-v1-MNRL                      & 0.380 & \textbf{0.440} & 0.418 & 0.408 & 0.400 & 0.396 & 0.394 & 0.392 \\
LangCache-Reranker-v2-MNRL                      & 0.380 & \textbf{0.436} & 0.403 & 0.386 & 0.374 & 0.368 & 0.365 & 0.363 \\
\midrule
\multicolumn{9}{c}{\textit{LangCache Rerankers (BCE)}} \\
LangCache-Reranker-v1-BCE                       & 0.378 & \textbf{0.435} & 0.402 & 0.382 & 0.363 & 0.352 & 0.343 & 0.337 \\
LangCache-Reranker-v2-BCE                       & 0.361 & \textbf{0.375} & 0.312 & 0.284 & 0.259 & 0.246 & 0.238 & 0.232 \\
\bottomrule
\end{tabular}
\caption{\textbf{P-CHR AUC vs.\ candidate pool size $K$} for all rerankers paired with LangCache-Embed-v3. Highest value per row in \textbf{bold}. The independently-scored cross-encoders (GTE-Rank, MiniLM, and the LangCache MNRL and BCE models) all peak at $K=2$ and decline as the pool grows; the ColBERT-family models score $0$ at $K=1$ and peak later, but only ColBERTv2.0 keeps improving toward $K=50$ while the other three peak by $K=5$--$10$ and then decline. At every $K$, only a model that beats the retriever baseline (top row) is worth deploying.}
\label{tab:k-sensitivity}
\end{table*}

\section{Overhead Latency}
\label{sec:latency}

Table~\ref{tab:latency} reports per-query latency with LangCache-Embed-v3 fixed as the retriever, so that reranking overhead is measured against a single retrieval cost. In absolute terms, the second stage is cheap: reranking adds between $17$ and $66$\,ms per query on the evaluation hardware, against a retrieval time of $26$\,ms, so total per-query latency stays under $100$\,ms for every configuration. Reranking accounts for $43$--$76\%$ of pipeline time, with the multi-vector ColBERT-family models the most expensive within that range. 

The cost is architectural rather than a matter of size: the $110$M-parameter ColBERTv2.0 is slower than the $150$M single-vector cross-encoders, because multi-vector scoring encodes every candidate separately rather than scoring a single query--candidate pair. Latency also does not buy deployment quality: MiniLM is the smallest and fastest model yet lands mid-pack on P-CHR AUC ($0.383$), while the most expensive ColBERT-family models rank no better, so compute spent on reranking does not translate into a better operating point. Against this retriever no reranker delivers positive deployment gain at all: the $\Delta$Ret column is negative for every model, least so for GTE-Rank ($-0.023$) and ColBERTv2.0 ($-0.040$) and worst for the BCE models ($-0.11$ to $-0.21$). Only three rerankers are undominated in the size--quality tradeoff: MiniLM as the smallest, GTE-Rank as the highest-scoring, and ColBERTv2.0, which reaches near-top quality at $110$M; every other model is matched or beaten by GTE-Rank at equal or smaller size.

These numbers are upper bounds and should be read as relative compute cost, not production latency. Latency grows with the candidate pool. We report metrics for the largest pool size $K=50$; smaller pools rerank fewer candidates at proportionally lower cost. The measurements also isolate the model forward pass, timed one query at a time against a co-located cache; they exclude the network, batching, concurrency, and serving overheads of a real deployment, and for the multi-vector models they re-encode candidates that a deployment would precompute. The practical cost of an unnecessary reranking stage is therefore better captured by the deployment-quality regressions of \S\ref{sec:results-calibration} than by wall-clock latency, which is small for every model we evaluate.

\begin{table*}[h]
\centering
\small
\begin{tabular}{lrrrrrr}
\toprule
Reranker & Size & P-CHR AUC & $\Delta$Ret & Mean (ms) & p95 (ms) & Overhead (\%) \\
\midrule
\multicolumn{7}{c}{\textit{ColBERT-family}} \\
ColBERTv2.0                  & 110M & 0.405 & $-0.040$ & 45 & 52 & 68.7 \\
Reason-ModernColBERT         & 149M & 0.384 & $-0.061$ & 64 & 71 & 75.6 \\
ColBERT-Zero                 & 149M & 0.379 & $-0.066$ & 66 & 74 & 72.1 \\
GTE-ModernColBERT-v1         & 149M & 0.357 & $-0.088$ & 64 & 72 & 75.4 \\
\midrule
\multicolumn{7}{c}{\textit{General cross-encoders}} \\
GTE-Reranker-ModernBERT-base & 150M & 0.422 & $-0.023$ & 41 & 52 & 64.2 \\
ms-marco-MiniLM-L12-v2       & 33M  & 0.383 & $-0.062$ & \textbf{17} & \textbf{22} & \textbf{43.1} \\
\midrule
\multicolumn{7}{c}{\textit{LangCache Rerankers (MNRL)}} \\
LangCache-Reranker-v1-MNRL                      & 150M & 0.392 & $-0.053$ & 41 & 52 & 64.0 \\
LangCache-Reranker-v2-MNRL                      & 150M & 0.363 & $-0.082$ & 41 & 52 & 64.0 \\
\midrule
\multicolumn{7}{c}{\textit{LangCache Rerankers (BCE)}} \\
LangCache-Reranker-v1-BCE                       & 150M & 0.337 & $-0.108$ & 42 & 53 & 62.1 \\
LangCache-Reranker-v2-BCE                       & 150M & 0.232 & $-0.213$ & 41 & 52 & 63.9 \\
\bottomrule
\end{tabular}
\caption{\textbf{Reranking latency per query} with LangCache-Embed-v3 as the retriever (retrieval $26$\,ms per query). Size is the total parameter count from each model's weights. P-CHR AUC is repeated from Table~\ref{tab:full-pchr} for reference; $\Delta$Ret is P-CHR AUC minus the retriever's P-CHR AUC ($0.445$), so a negative value means reranking lowers deployment quality below the retriever alone. Mean and p95 are reranking-only forward-pass times at $K=50$; Overhead is reranking as a fraction of total pipeline time; lowest-latency model in \textbf{bold}. Measured in isolation on the evaluation hardware, one query at a time; not end-to-end production latency.}
\label{tab:latency}
\end{table*}

\section{Score Distribution Analysis}
\label{sec:distribution-plots}

To explain \textit{why} threshold-utility gaps arise, we analyze the score distributions of $c^*$ separately, comparing positive queries ($y=1$) against negative queries ($y=0$). A well-placed model assigns systematically higher scores to positives, with minimal shared probability mass near the decision boundary. We use three separation metrics, and additionally report the standard probability-calibration metrics (ECE, NLL, Brier) to test whether they explain deployment.

\paragraph{ROC-AUC.} The area under the receiver operating characteristic curve is the probability that a random positive query scores above a random negative one. It is the standard threshold-independent summary of rank separation, ranging from 0.5 (chance) to 1.0 (perfect ranking). Like PR-AUC, it is invariant to monotonic rescaling and therefore blind to \emph{where} on the scale the scores sit.

\paragraph{KS Statistic.} The Kolmogorov--Smirnov statistic measures the maximum absolute distance between the cumulative distribution functions (CDFs) of the positive and negative score distributions:
\begin{equation}
  \text{KS} = \sup_{s} \left| F_{\text{pos}}(s) - F_{\text{neg}}(s) \right|
\end{equation}

where $F_{\text{pos}}$ and $F_{\text{neg}}$ are the empirical CDFs of $s(q, c^*)$ for positive and negative queries respectively. KS ranges from 0 (identical distributions) to 1 (perfect separation), and is equivalent to the maximum achievable gap between true positive rate and false positive rate across all thresholds.

\paragraph{Distribution Overlap.} While KS captures the single best threshold, it can mask ambiguity elsewhere in the score range. We therefore also measure the total shared probability mass between the two score distributions. Let $f_{\text{pos}}(s)$ and $f_{\text{neg}}(s)$ denote their probability density functions; since all scores lie in $[0,1]$ after normalization, overlap is:
\begin{equation}
  \text{Overlap} = \int_0^1 \min\!\bigl(f_{\text{pos}}(s),\, f_{\text{neg}}(s)\bigr)\, ds
\end{equation}
In practice, $f_{\text{pos}}$ and $f_{\text{neg}}$ are estimated via Gaussian kernel density estimation (KDE) from empirical scores; overlap is then computed numerically over the score range using the trapezoidal rule. Lower overlap indicates less ambiguity across the full score range.

Table~\ref{tab:distribution-metrics} reports these metrics for all rerankers paired with LangCache-Embed-v3, alongside their exact P-CHR AUC for that retriever. None predicts P-CHR AUC; the relationship is at best weak and at the extremes inverted: the four most-overlapping models (the ColBERT family) deploy among the best, while the least-overlapping model (LangCache-Reranker-v1-BCE) deploys near the bottom.

\paragraph{ColBERT-family models are the clearest contradiction.} All four are the worst-separated rerankers, with the lowest ROC-AUC and KS statistics and the highest overlap. Yet their P-CHR AUC sits well above both BCE rerankers and is competitive with the MNRL pair; ColBERTv2.0 (0.405) trails only the GTE cross-encoder (0.422) across all ten. As established in \S\ref{sec:results-colbert}, softmax normalization over the candidate pool gives ColBERT scores a natural spread, so the overlap that the KDE metric penalizes still leaves precision-favorable operating points available.

\paragraph{Best separation, poor deployment.} At the opposite extreme, LangCache-Reranker-v1-BCE has the lowest overlap of any reranker (0.231) and strong ranking scores (ROC-AUC 0.870, KS 0.612), yet its P-CHR AUC is only 0.337, below the retriever and every ColBERT model. The failure lies in where the overlap is located, not how much there is: BCE training concentrates negative scores near 0.0, directly overlapping the lower tail of the positive distribution. Any threshold low enough to admit weakly positive queries simultaneously fires on this dense negative mass, collapsing precision.

\paragraph{Training objective determines score placement, not scale.} MNRL outperforms BCE in deployment at both scales (v1: 0.392 vs.\ 0.337; v2: 0.363 vs.\ 0.232), and the separation metrics do not reliably track the gap. For the v1 pair they actively mislead, ranking BCE ahead of MNRL on overlap even though MNRL deploys better. MNRL distributes scores across a broader usable range, supplying viable operating points across many thresholds, whereas BCE compresses scores regardless of how the standard separation statistics happen to read. Training objective, not dataset size, is the decisive factor.

\begin{table*}[t]
\centering
\small
\begin{tabular}{lrrrrrrr}
\toprule
Reranker & ROC-AUC & KS & Overlap & ECE & NLL & Brier & P-CHR AUC \\
\midrule
\multicolumn{8}{c}{\textit{ColBERT-family}} \\
ColBERTv2.0                    & 0.610 & 0.202 & 0.634 & 0.324 & 2.39 & 0.361 & 0.405 \\
Reason-ModernColBERT           & 0.622 & 0.203 & 0.749 & 0.306 & 1.34 & 0.339 & 0.384 \\
ColBERT-Zero                   & 0.622 & 0.203 & 0.778 & 0.340 & 1.39 & 0.363 & 0.379 \\
GTE-ModernColBERT-v1           & 0.625 & 0.202 & 0.782 & 0.415 & 1.73 & 0.416 & 0.357 \\
\midrule
\multicolumn{8}{c}{\textit{General cross-encoders}} \\
GTE-Reranker-ModernBERT-base   & 0.813 & 0.525 & 0.415 & 0.355 & 1.26 & 0.334 & \textbf{0.422} \\
ms-marco-MiniLM-L12-v2         & 0.713 & 0.381 & 0.400 & 0.287 & 2.08 & 0.300 & 0.383 \\
\midrule
\multicolumn{8}{c}{\textit{LangCache Rerankers (MNRL)}} \\
LangCache-Reranker-v1-MNRL                        & \textbf{0.888} & \textbf{0.653} & 0.328 & 0.081 & \textbf{0.64} & \textbf{0.137} & 0.392 \\
LangCache-Reranker-v2-MNRL                        & 0.872 & 0.616 & 0.372 & \textbf{0.077} & 0.65 & 0.147 & 0.363 \\
\midrule
\multicolumn{8}{c}{\textit{LangCache Rerankers (BCE)}} \\
LangCache-Reranker-v1-BCE                         & 0.870 & 0.612 & \textbf{0.231} & 0.176 & 1.51 & 0.181 & 0.337 \\
LangCache-Reranker-v2-BCE                         & 0.839 & 0.563 & 0.422 & 0.343 & 1.12 & 0.321 & 0.232 \\
\bottomrule
\end{tabular}
\caption{\textbf{Score-distribution and probability-calibration metrics.} ROC-AUC, KS, overlap, ECE, NLL, Brier, and exact P-CHR AUC for all 10 rerankers paired with LangCache-Embed-v3 (retriever baseline P-CHR AUC $= 0.445$). Best value per metric in \textbf{bold} (lowest for overlap, ECE, NLL, Brier). Neither the separation metrics (ROC-AUC, KS, overlap) nor the probability-calibration metrics (ECE, NLL, Brier) predict deployment: across all 90 combinations ECE and NLL correlate only weakly with exact P-CHR AUC (Spearman $\rho = 0.22$ and $0.27$) and Brier moderately ($0.50$), the last because it conflates calibration with discrimination. ColBERT-family models are the worst-separated and among the worst-calibrated (highest ECE) yet competitive on P-CHR AUC.}
\label{tab:distribution-metrics}
\end{table*}

\begin{figure*}[t]
\centering
\begin{subfigure}[t]{0.32\textwidth}
\centering
\includegraphics[height=2.25cm]{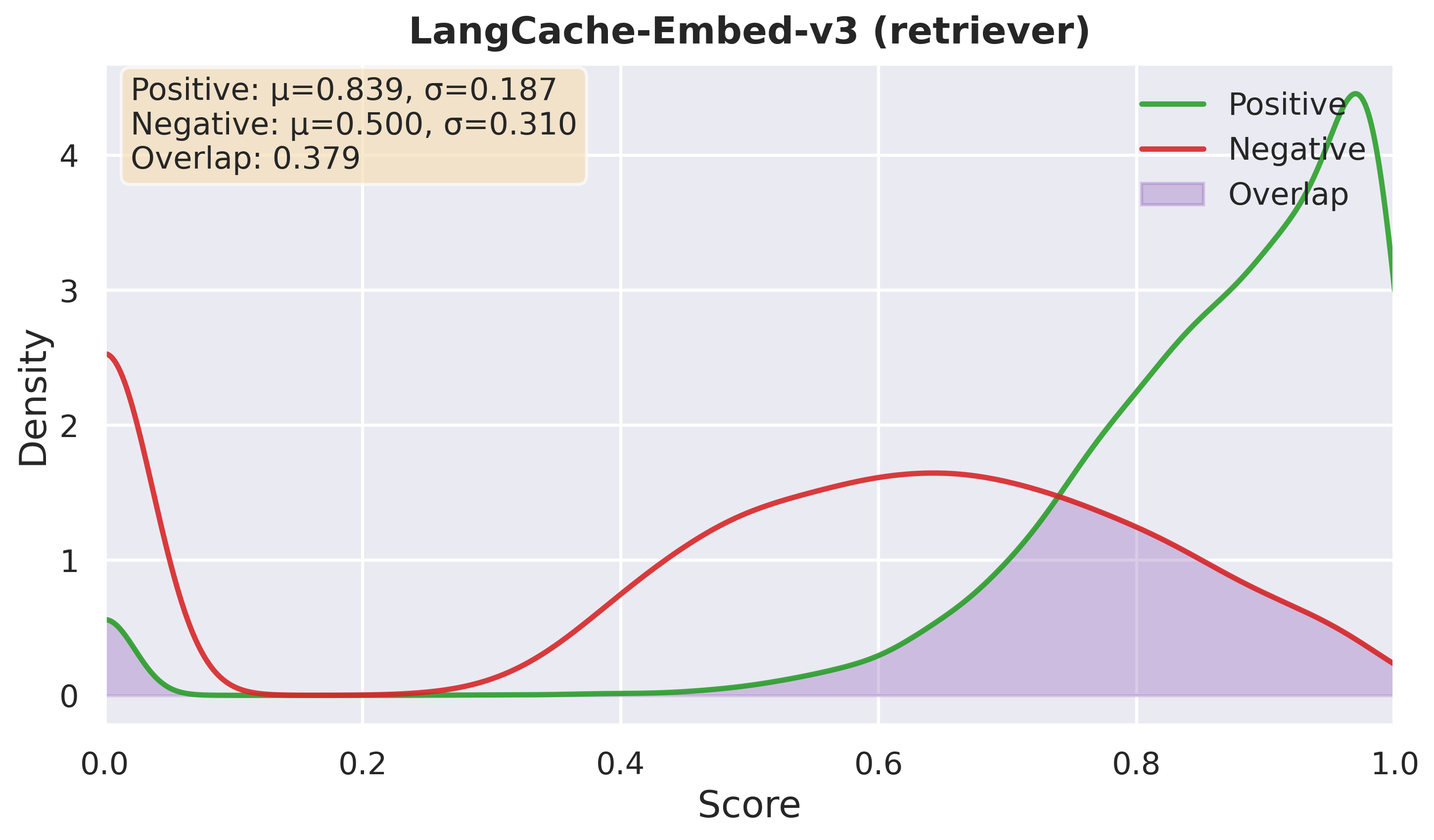}
\caption{\textbf{Retriever (LangCache-Embed-v3).} Positives clean at $1.0$, negatives centred at $0.5$; the overlap sits in the upper-mid range, leaving the top of the scale usable, giving the best P-CHR AUC ($0.445$).}
\end{subfigure}
\hfill
\begin{subfigure}[t]{0.32\textwidth}
\centering
\includegraphics[height=2.25cm]{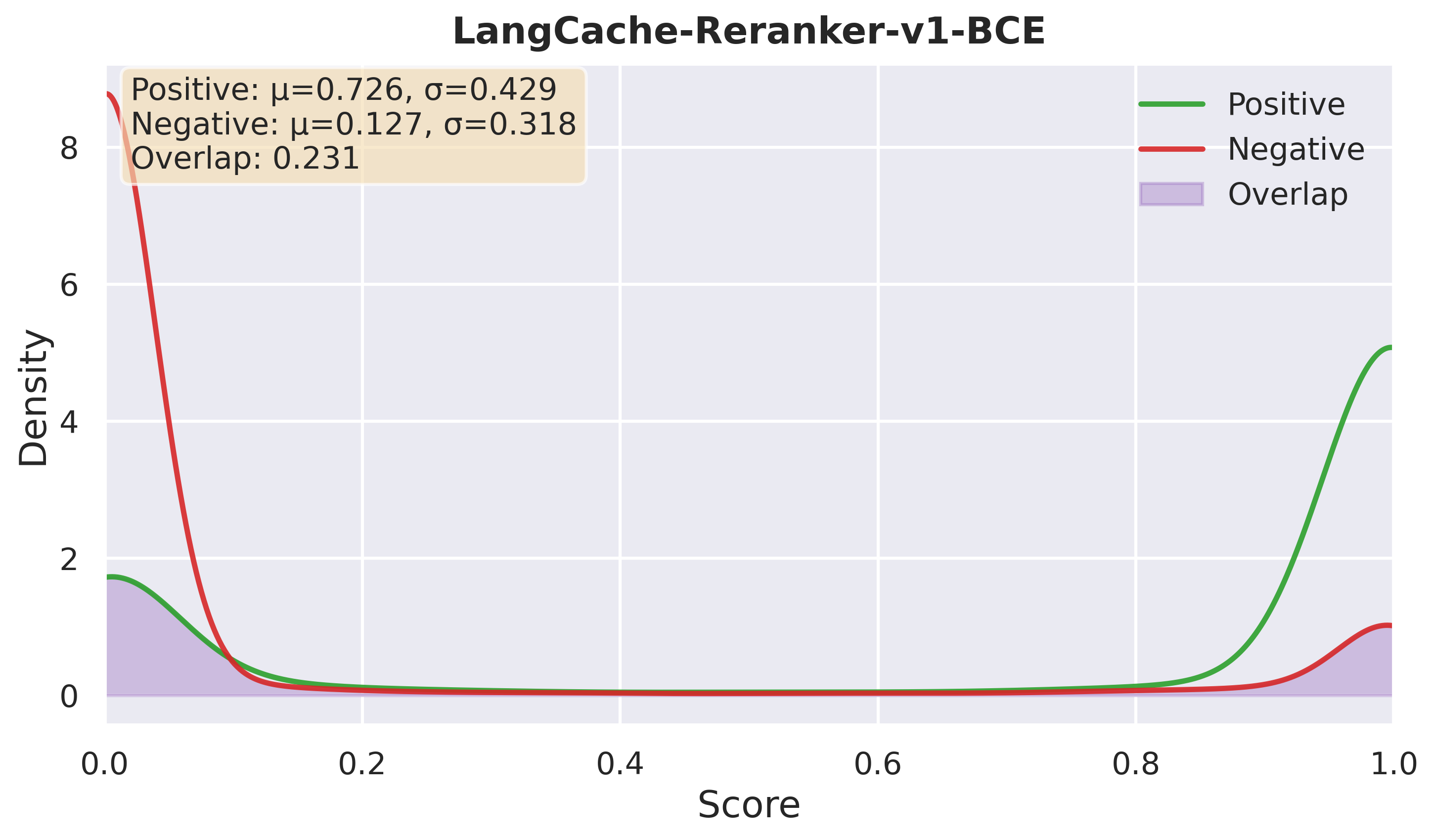}
\caption{\textbf{LangCache-Reranker-v1-BCE: boundary collapse.} Negatives mass at $0.0$ overlapping the positive tail; the lowest overlap ($0.231$) sits at the operating zone, depressing P-CHR AUC to $0.337$, below the retriever and every ColBERT model.}
\end{subfigure}
\hfill
\begin{subfigure}[t]{0.32\textwidth}
\centering
\includegraphics[height=2.25cm]{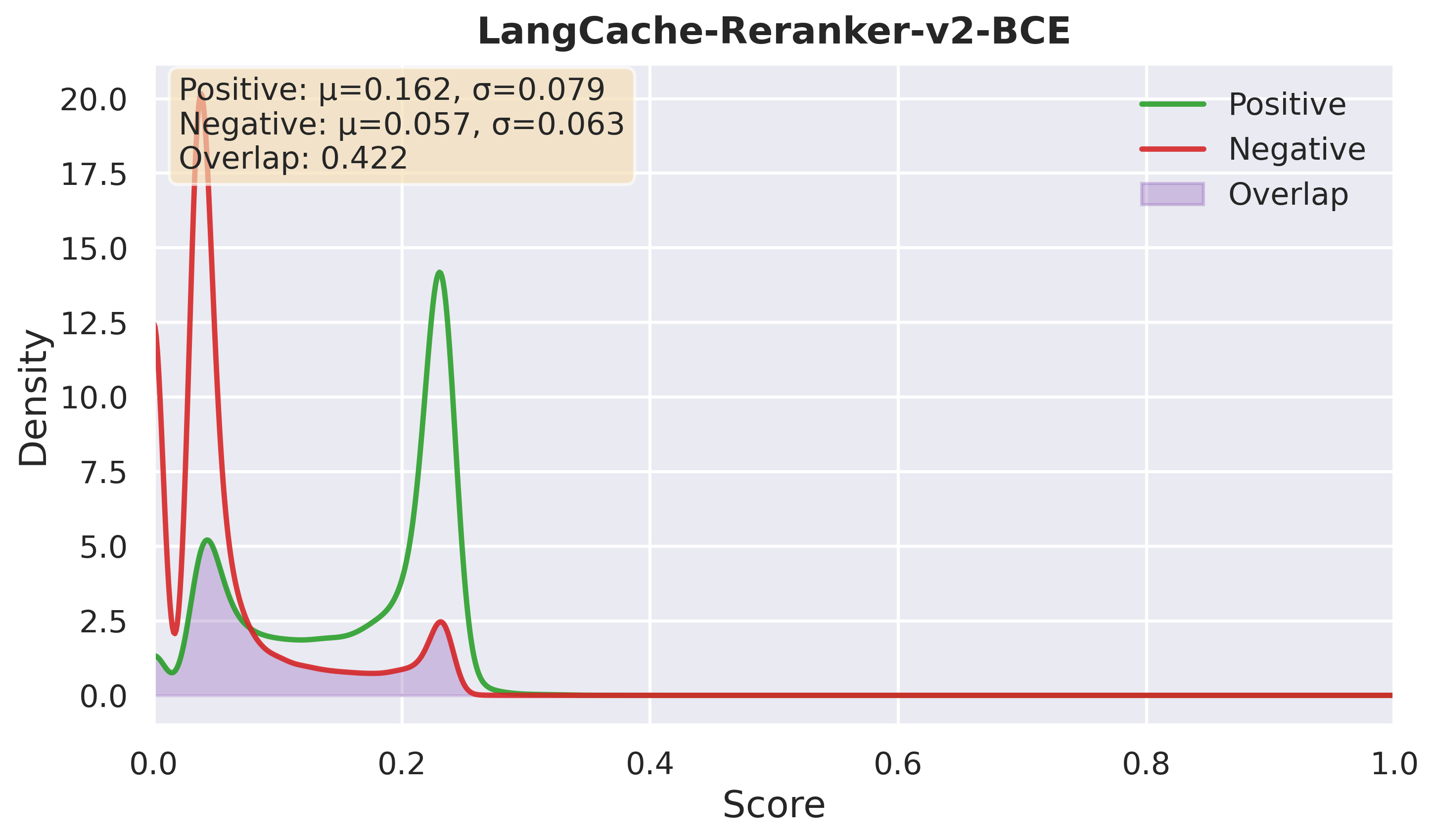}
\caption{\textbf{LangCache-Reranker-v2-BCE: score compression.} All mass in $[0,0.3]$ with ordering preserved but few distinct thresholds; higher overlap ($0.422$) yet the worst P-CHR AUC of any reranker ($0.232$).}
\end{subfigure}

\medskip
\begin{subfigure}[t]{0.32\textwidth}
\centering
\includegraphics[height=2.25cm]{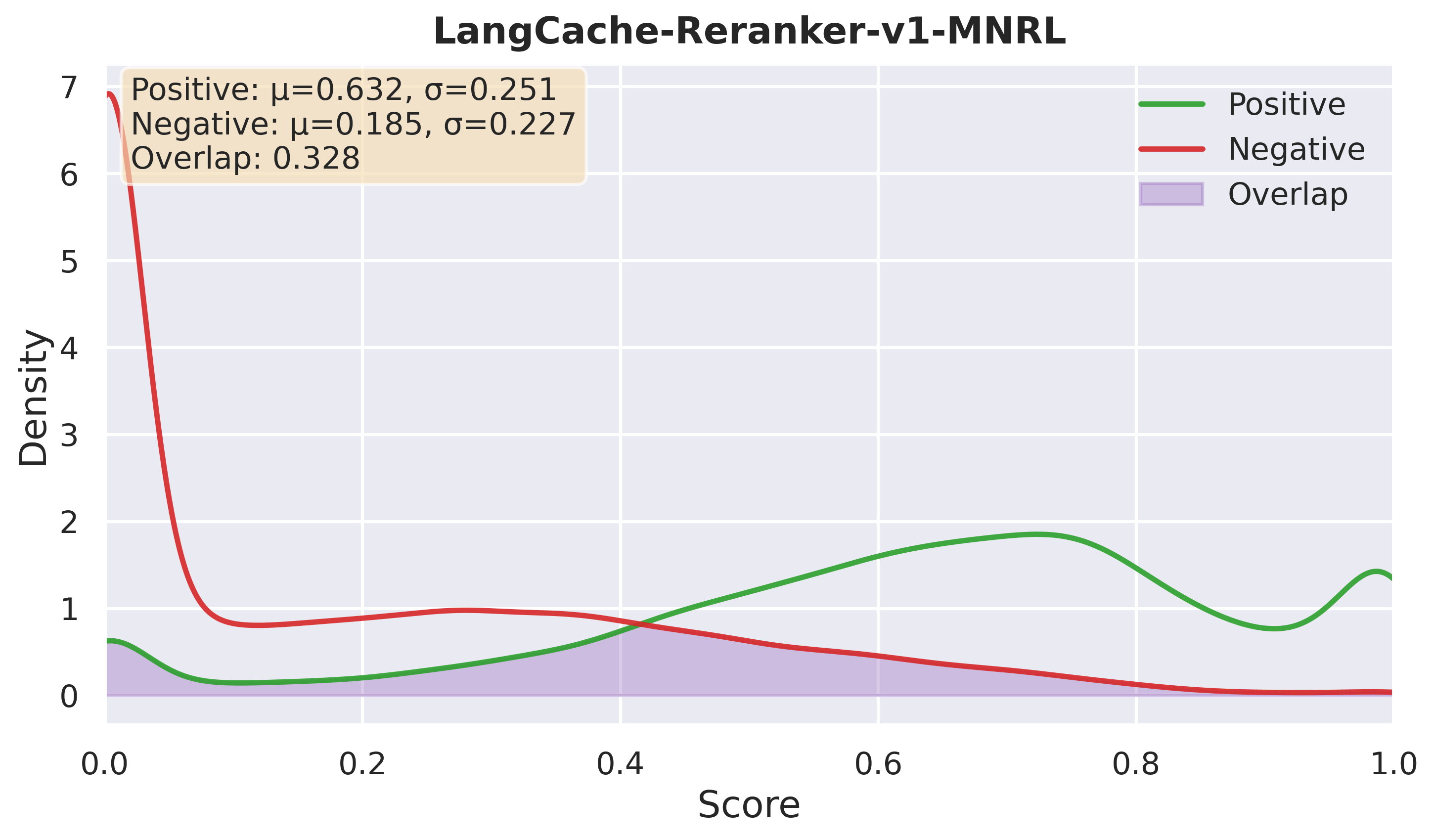}
\caption{\textbf{LangCache-Reranker-v1-MNRL: graded spread.} Broad positive hump with negatives held at $0.0$ (overlap $0.328$) supplies many precision-favorable thresholds; P-CHR AUC $0.392$.}
\end{subfigure}
\hfill
\begin{subfigure}[t]{0.32\textwidth}
\centering
\includegraphics[height=2.25cm]{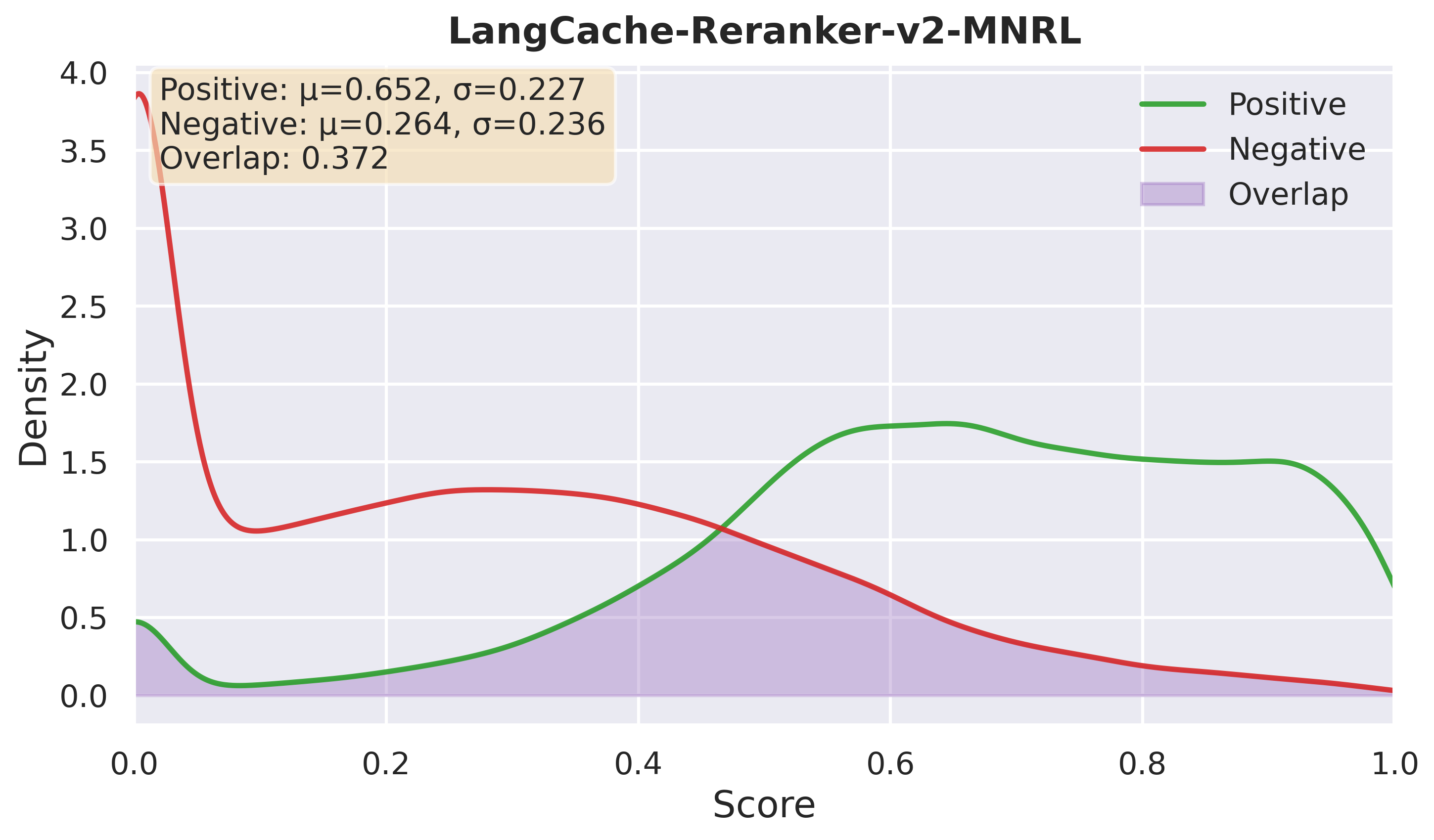}
\caption{\textbf{LangCache-Reranker-v2-MNRL: graded spread.} Same shape at 40M pairs but negatives spread into the mid-range (overlap $0.372$); slightly lower P-CHR AUC ($0.363$).}
\end{subfigure}
\hfill
\begin{subfigure}[t]{0.32\textwidth}
\centering
\includegraphics[height=2.25cm]{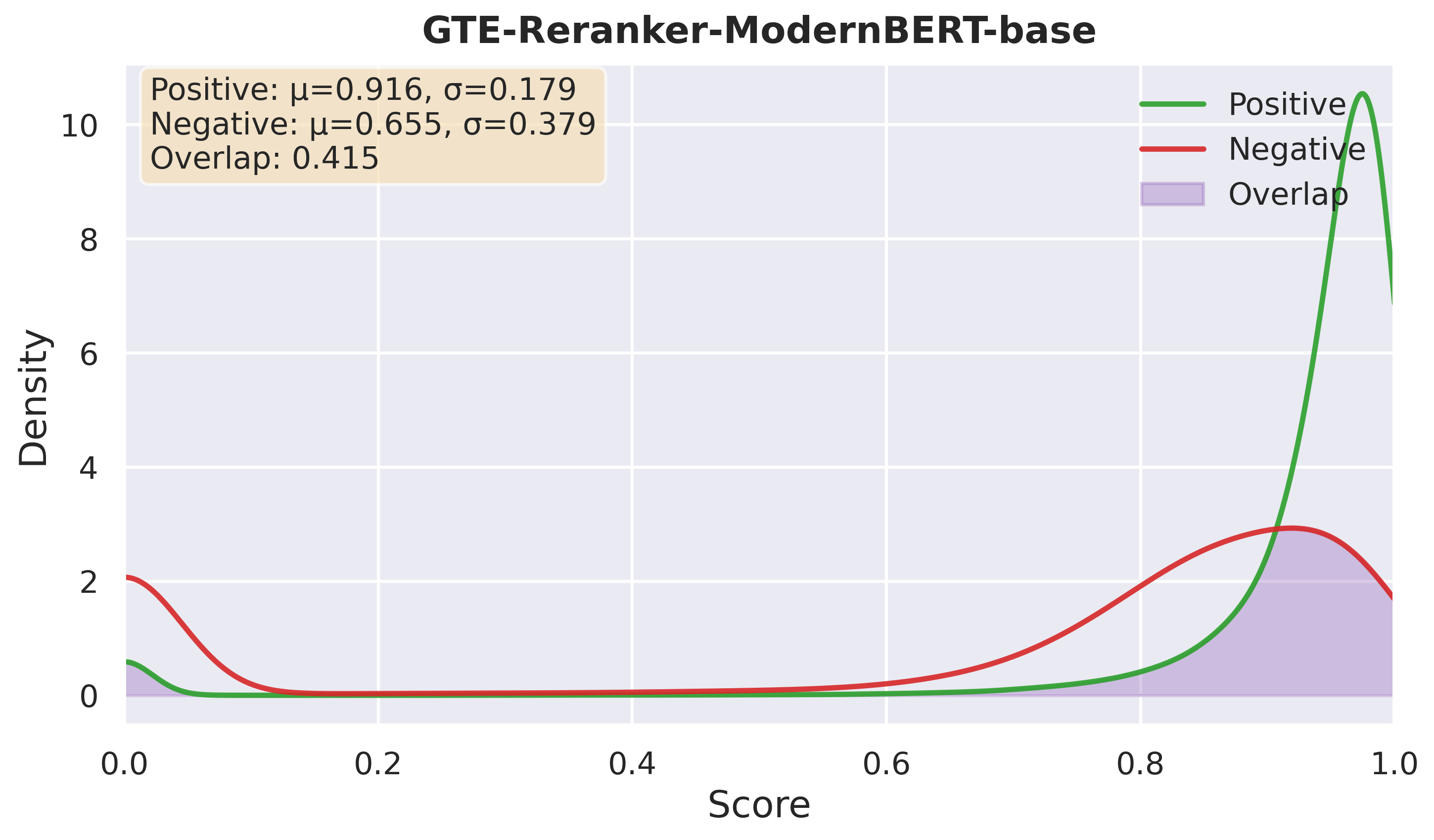}
\caption{\textbf{GTE-Reranker: high-score FPs.} A negative hump near $1.0$ forms a false-positive floor, but the positive spike still dominates the top, giving the best reranker P-CHR AUC ($0.422$).}
\end{subfigure}

\medskip
\begin{subfigure}[t]{0.32\textwidth}
\centering
\includegraphics[height=2.25cm]{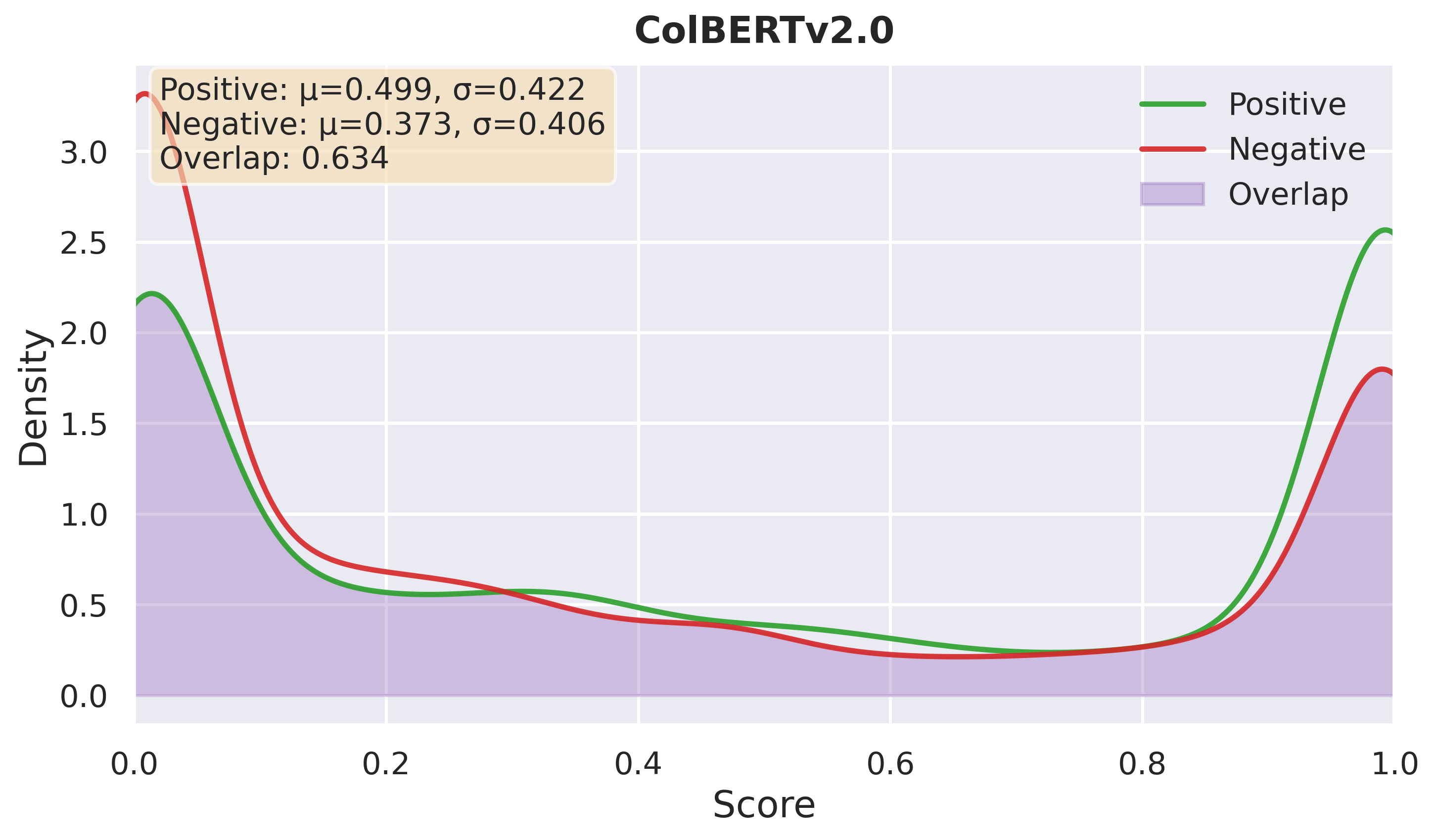}
\caption{\textbf{ColBERTv2.0: softmax spread.} Scores span the full range and positives dominate the top despite high overlap ($0.634$); P-CHR AUC $0.405$.}
\end{subfigure}
\hfill
\begin{subfigure}[t]{0.32\textwidth}
\centering
\includegraphics[height=2.25cm]{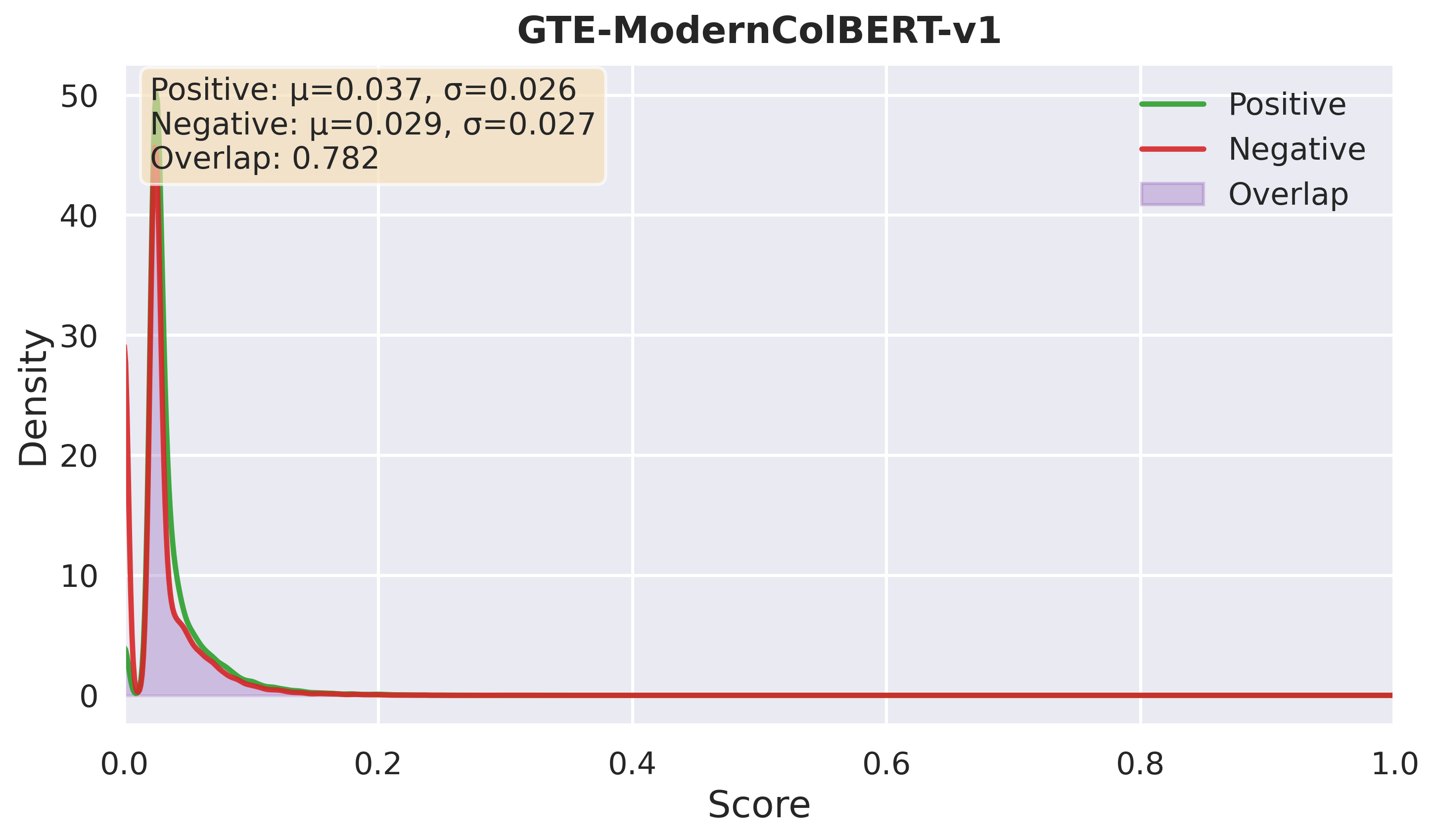}
\caption{\textbf{GTE-ModernColBERT-v1: softmax spread.} Scores compressed near $0.0$ with the highest overlap ($0.782$), yet per-query ordering still deploys above BCE; P-CHR AUC $0.357$.}
\end{subfigure}
\hfill
\begin{subfigure}[t]{0.32\textwidth}
\centering
\includegraphics[height=2.25cm]{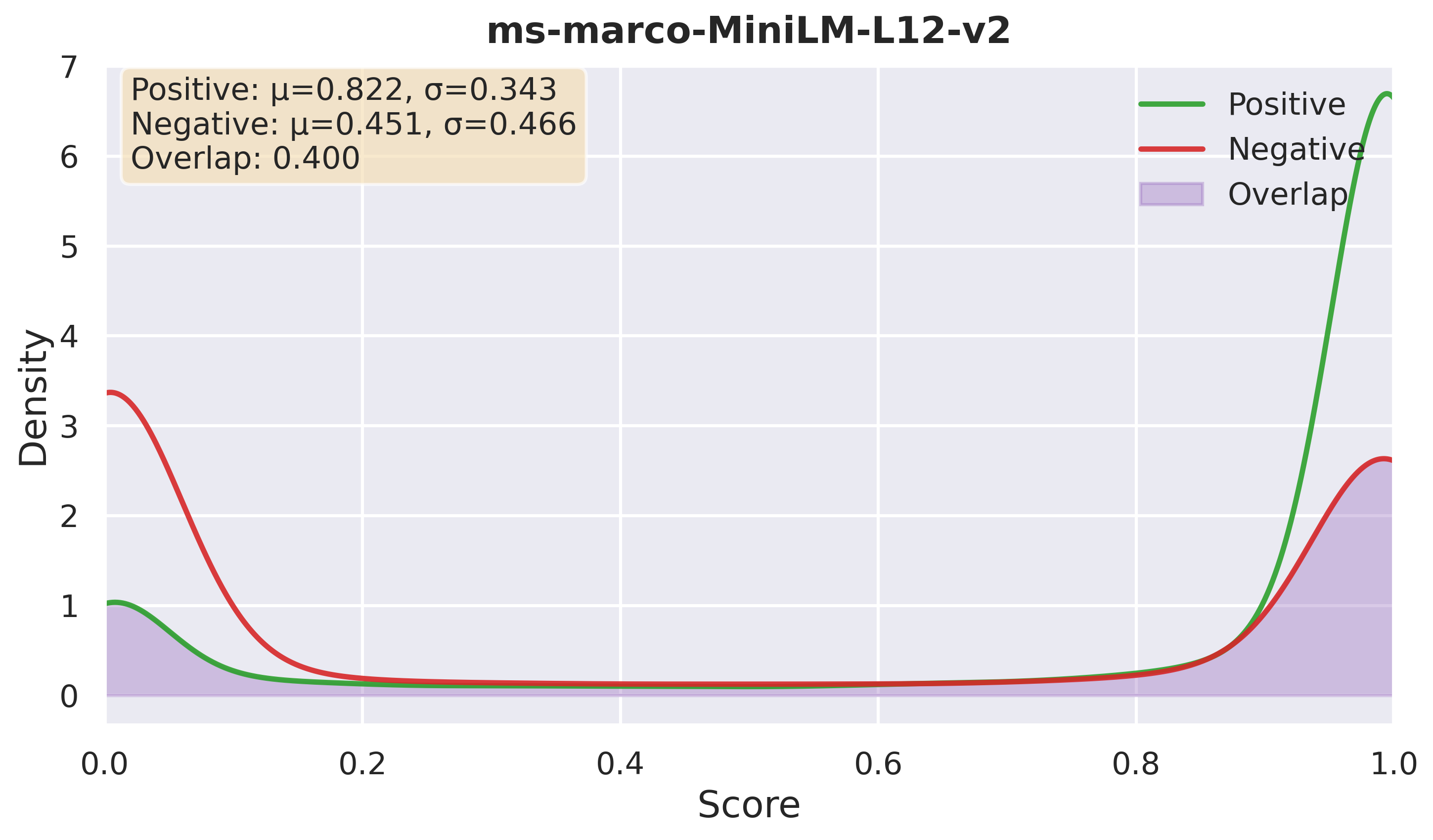}
\caption{\textbf{MiniLM: high-score FPs.} Large negative mass at $1.0$ under the positive spike; a more severe false-positive floor than GTE-Reranker, giving P-CHR AUC $0.383$.}
\end{subfigure}
\caption{\textbf{KDE of ground-truth scores $s(q,c^*)$ by label} (positive: green, negative: red, overlap: purple) for the retriever and eight rerankers paired with LangCache-Embed-v3 at $K=50$. Deployment quality is set by where the score mass sits relative to a usable threshold, not by aggregate overlap: LangCache-Reranker-v1-BCE has the lowest overlap yet a P-CHR AUC below the retriever and every ColBERT model, while the ColBERT-family panels have the highest overlap yet deploy above both BCE models.}
\label{fig:dist-kde}
\end{figure*}

This section presents per-reranker kernel density estimates of the ground-truth score $s(q,c^*)$, separated by label, for nine configurations paired with LangCache-Embed-v3 (Figure~\ref{fig:dist-kde}; a representative subset of Table~\ref{tab:distribution-metrics}). The plots make the deployment mechanism visible: what governs P-CHR AUC is \emph{where} each model places its score mass relative to a usable threshold, not how much the two distributions overlap. Five distinct mechanisms emerge.

\paragraph{Retriever baseline.} The retriever concentrates positives near $1.0$ ($\mu_{\text{pos}}=0.84$) while negatives form a broad hump centred on $0.5$ ($\mu_{\text{neg}}=0.50$); the overlap sits in the upper-mid range and leaves the top of the scale clean. This placement, not any reranker's, gives the best P-CHR AUC ($0.445$).

\paragraph{Boundary collapse (LangCache-Reranker-v1-BCE).} BCE concentrates negatives at $0.0$ ($\mu_{\text{neg}}=0.13$) directly overlapping the positive lower tail, while the remaining positives spike at $1.0$ ($\mu_{\text{pos}}=0.73$). This produces the lowest overlap of any model ($0.231$) but places it exactly where a low threshold must operate, so any threshold admitting weak positives also fires on the dense negative mass and P-CHR AUC falls to $0.337$, below the retriever and every ColBERT model.

\paragraph{Score compression (LangCache-Reranker-v2-BCE).} Scaling BCE to 40M pairs pulls all mass into $[0,0.3]$, with positives peaking at $0.23$ above negatives at $0.04$ ($\mu_{\text{pos}}=0.16$, $\mu_{\text{neg}}=0.06$). Because the positive ordering is preserved within this band it still deploys a nonzero P-CHR AUC ($0.232$), but the compressed range exposes few distinct operating points, making it the worst reranker overall, now below LangCache-Reranker-v1-BCE's $0.337$ despite its own higher overlap ($0.422$).

\paragraph{Graded spread (MNRL).} Both MNRL models place positives in a broad hump over the upper-mid range ($\mu_{\text{pos}}=0.63$--$0.65$) with negatives concentrated low, crossing around $0.45$ and supplying many precision-favorable thresholds. LangCache-Reranker-v1-MNRL keeps negatives sharply at $0.0$ (overlap $0.328$) whereas the 40M LangCache-Reranker-v2-MNRL spreads them into the mid-range (overlap $0.372$), and LangCache-Reranker-v1-MNRL deploys slightly better ($0.392$ vs.\ $0.363$): the objective fixes the favorable shape and scale only modulates it.

\paragraph{High-score false positives (GTE-Reranker, MiniLM).} The general cross-encoders push positives hard to $1.0$ but also raise a negative hump near the top ($\mu_{\text{neg}}=0.66$ for GTE-Reranker), creating a false-positive floor. GTE-Reranker deploys best of all rerankers ($0.422$) because its positive spike at $1.0$ still dominates the very top; MiniLM places more negative mass at $1.0$ and sits lower at $0.383$.

\paragraph{Softmax spread (ColBERT-family).} Softmax normalization over the candidate pool gives ColBERT the highest overlaps: ColBERTv2.0 spreads scores across the whole range (overlap $0.634$) while GTE-ModernColBERT-v1 compresses them near $0.0$ (overlap $0.782$). Yet, both keep positives above negatives within each query's pool, so precision-favorable thresholds remain. Both deploy above either BCE model despite the worst separation statistics, confirming the effect is a property of the family's normalization rather than of ColBERTv2.0 alone.

\section{Performance Curves}
\label{sec:curves}

\begin{figure*}[t]
\centering
\begin{subfigure}[t]{0.49\textwidth}
\centering
\includegraphics[width=\linewidth]{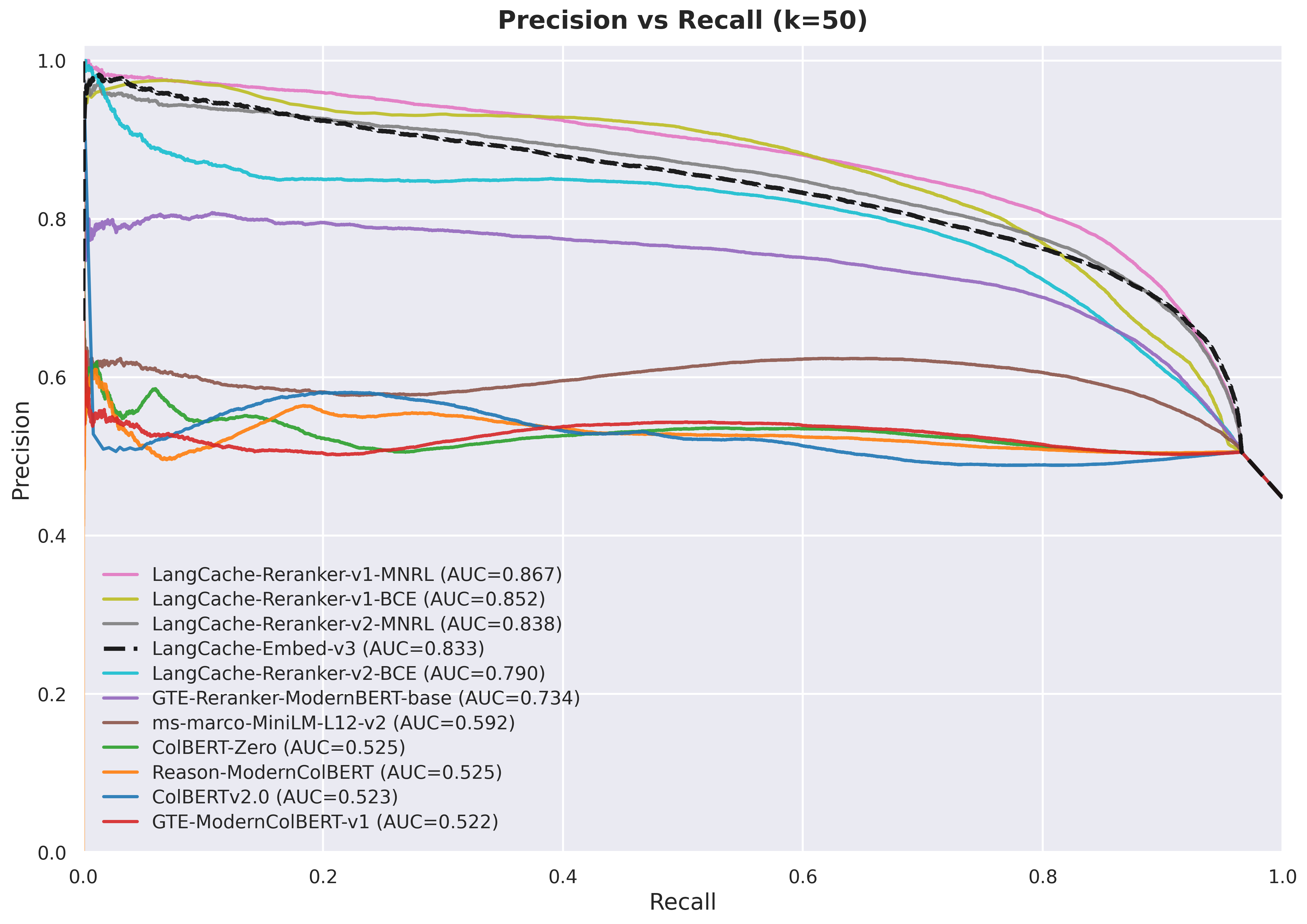}
\caption{\textbf{Precision--Recall.} Every model looks healthy; BCE rerankers rank near the top.}
\end{subfigure}
\hfill
\begin{subfigure}[t]{0.49\textwidth}
\centering
\includegraphics[width=\linewidth]{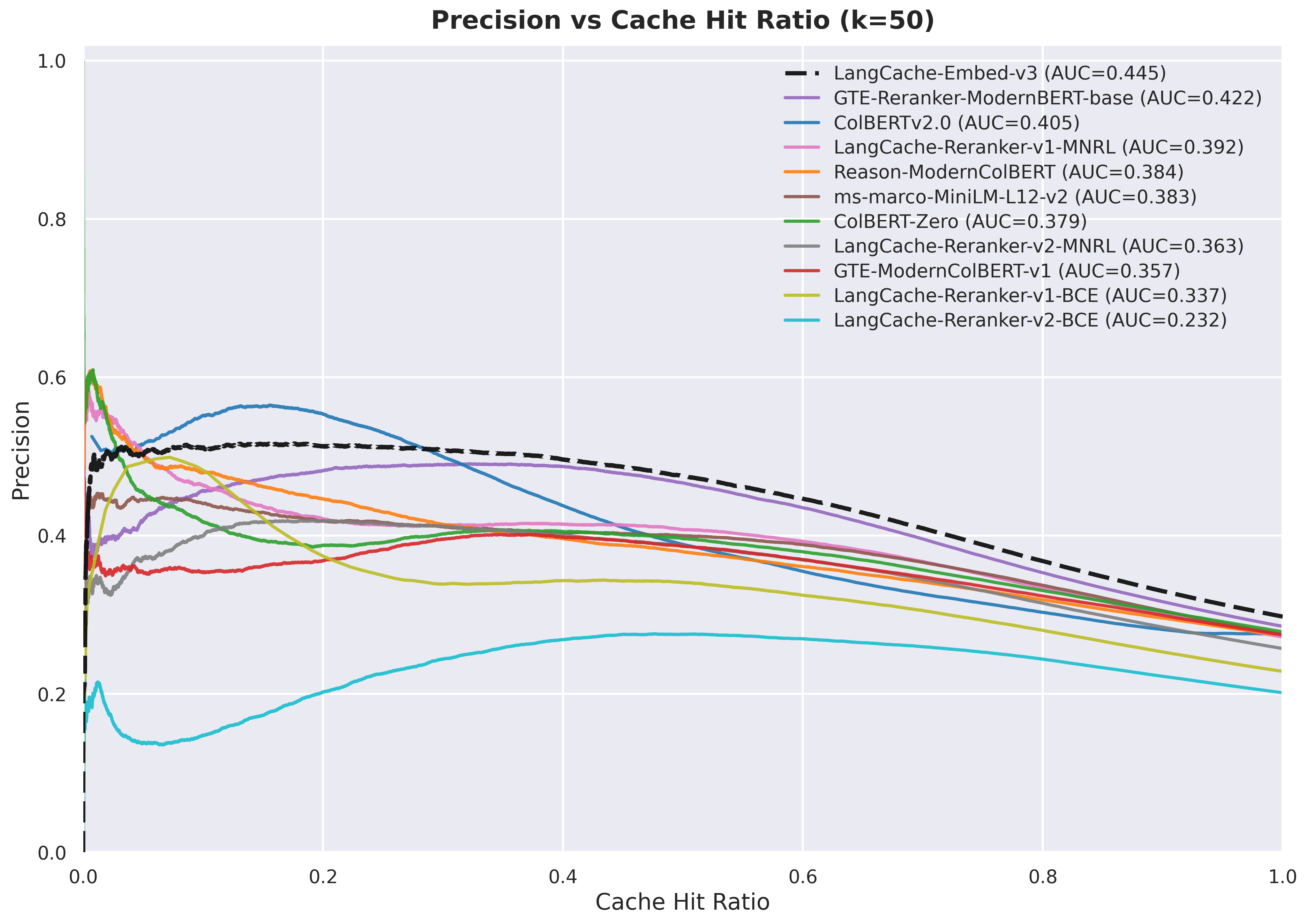}
\caption{\textbf{Precision--CHR.} BCE collapses; ColBERT-family climbs; no reranker beats the retriever.}
\end{subfigure}

\caption{\textbf{Offline vs.\ deployment curves} for all ten rerankers paired with LangCache-Embed-v3 at $K=50$ (retriever baseline as a bold black dashed line; legends sorted by AUC). The PR view (a) ranks the BCE rerankers near the best and the ColBERT-family worst; the P-CHR view (b) inverts this, exposing the BCE precision collapse that PR-AUC hides.}
\label{fig:curves}
\end{figure*}

This section contrasts the Precision--Recall and Precision--CHR curves for all ten rerankers paired with LangCache-Embed-v3 at $K=50$ (Figure~\ref{fig:curves}). The gap between the offline and deployment views is the central diagnostic of this work: the PR curves look healthy for every model, whereas the corresponding P-CHR curves expose the precision collapse of the BCE rerankers as $\tau$ is lowered to raise cache utilization.

\paragraph{PR curves hide threshold-utility failures.} On the Precision--Recall axes (Figure~\ref{fig:curves}a) every configuration traces a smooth, well-behaved curve, with the LangCache-Reranker-v1-BCE reranker being among the strongest (PR-AUC $0.852$); it sits second only to LangCache-Reranker-v1-MNRL ($0.867$) and closely tracks it, while the ColBERT-family models rank lowest ($\approx 0.52$). Nothing in this view signals a deployment problem; selected by PR-AUC, the BCE rerankers look like top choices.

\paragraph{P-CHR curves expose the deployment hierarchy.} Re-plotting the same models against cache utilization (Figure~\ref{fig:curves}b) reshuffles the ranking and fans the curves apart. The retriever baseline (bold black dashed line, P-CHR AUC $0.445$) has the highest area, and no reranker stays above it as utilization rises, so reranking does not improve deployment on this retriever. The PR-laggard ColBERT-family becomes competitive (ColBERTv2.0 $0.405$, behind the best reranker, GTE-Reranker at $0.422$), whereas the BCE rerankers collapse to a low band, their precision suppressed at every utilization level despite their high PR-AUC. All curves fall as CHR${\to}1$, toward the low precision of a cache that fires on nearly every query. The inversion is stark: LangCache-Reranker-v1-BCE falls from second overall on PR to the bottom band on P-CHR, and the ColBERT-family climbs from the bottom of the PR ranking to the front of the rerankers.

Restricting to valid fires (P-VCHR) preserves the same ordering, confirming the hierarchy is not an artifact of how cache fires are tallied; per-combination P-VCHR AUC values are reported in Table~\ref{tab:full-pvchr}.

\section{Score-Normalization Ablation}
\label{sec:normalization-ablation}

\begin{table}[!h]
\centering
\small
\setlength{\tabcolsep}{4pt}
\resizebox{\columnwidth}{!}{%
\begin{tabular}{llrr}
\toprule
Reranker & Alt.\ transform & Native & Alt. \\
\midrule
\multicolumn{4}{c}{\textit{ColBERT-family (native: pool-softmax)}} \\
ColBERTv2.0          & sigmoid      & 0.403 & 0.020 \\
Reason-ModernColBERT & sigmoid      & 0.382 & 0.196 \\
ColBERT-Zero         & sigmoid      & 0.375 & 0.167 \\
GTE-ModernColBERT-v1 & sigmoid      & 0.354 & 0.160 \\
\midrule
\multicolumn{4}{c}{\textit{Cross-encoders \& LangCache (native: sigmoid)}} \\
GTE-Reranker-ModernBERT-base & pool-softmax & 0.390 & 0.446 \\
ms-marco-MiniLM-L12-v2       & pool-softmax & 0.346 & 0.395 \\
LangCache-Reranker-v1-MNRL   & pool-softmax & 0.353 & 0.367 \\
LangCache-Reranker-v2-MNRL   & pool-softmax & 0.330 & 0.377 \\
LangCache-Reranker-v1-BCE    & pool-softmax & 0.290 & 0.344 \\
LangCache-Reranker-v2-BCE    & pool-softmax & 0.171 & 0.267 \\
\bottomrule
\end{tabular}%
}
\caption{\textbf{Score-normalization ablation.} Exact P-CHR AUC (averaged over 9 retrievers) under each model's native transform and the counterfactual one. Stripping the pool-softmax collapses the ColBERT family; applying it to the cross-encoders recovers much of the gap for the compressed (BCE) models but little for the already-spread MNRL models.}
\label{tab:normalization-ablation}
\end{table}

To separate ColBERT's representation from its pool-softmax normalization, we cross the two score transforms against the two model families: we apply ColBERT's pool-softmax to the cross-encoders (replacing their per-pair sigmoid) and strip it from the ColBERT family (scoring raw MaxSim with a per-pair sigmoid). Table~\ref{tab:normalization-ablation} reports exact P-CHR AUC, averaged over the 9 retrievers, for each model under its native transform and under the counterfactual one it does not natively use.

Stripping the pool-softmax collapses the ColBERT family (ColBERTv2.0 falls from $0.403$ to $0.020$, below every model), so the family's deployment advantage is a property of the pool-relative normalization, not of the MaxSim representation. Applied in the other direction, the same transform recovers much of the gap for the poorly-placed cross-encoders: LangCache-Reranker-v2-BCE gains $+0.096$, v1-BCE $+0.054$, and ms-marco-MiniLM-L12-v2 $+0.049$. It does little for the MNRL rerankers ($+0.013$ for v1-MNRL), whose scores are already well spread. Pool-softmax over the candidate pool is therefore a training-free deployment fix for models with compressed scores and a near-no-op for those already well-placed: a concrete remedy beyond changing the training objective, and one that cannot be achieved by post-hoc monotone calibration (Appendix~\ref{sec:calibration-fitting}).

\section{Practitioners' Guide}
\label{sec:extended-discussion}

\paragraph{Training objective, not scale, governs threshold utility.} The training objective, not dataset scale, decides whether a model's scores are usable at a threshold. Scaling the data 38$\times$ from the v1 to the v2 reranker generation left both objectives essentially unchanged in deployment (\S\ref{sec:results-calibration}), yet MNRL retained substantially more offline quality than BCE at either scale, nearly twice as much at the larger v2 scale (ORR $0.410$ vs.\ $0.229$). The cause is score semantics, not training signal: BCE collapses most mass at its decision boundary, whereas MNRL and multi-vector MaxSim models spread scores across a usable range. Larger paraphrase corpora cannot fix the wrong objective.

\paragraph{The operational gap has a floor no rescaling can cross.} The structural component of the operational gap (\S\ref{sec:metrics-gap}) is fixed by the positive rate and bounds P-CHR AUC for every model. The remainder does not yield to post-hoc calibration either: because P-CHR AUC is rank-invariant, temperature and Platt scaling leave it unchanged (Appendix~\ref{sec:calibration-fitting}). What recovers it is re-normalizing the score over the candidate pool (pool-softmax; Appendix~\ref{sec:normalization-ablation}) or choosing a better-placed objective, not a monotone knob. Because the floor moves with the positive rate, absolute P-CHR AUC does not transfer across deployments; ORR, normalized by ranking quality, is the more portable comparison.

\paragraph{Reranking is not free, and often not worth it.} Most rerankers lowered deployment precision below the retriever alone, and only ColBERTv2.0 and GTE-Reranker beat the average baseline (\S\ref{sec:results-calibration}); even they fell short of LangCache-Embed-v3 with no reranker. A stage that adds latency (Appendix~\ref{sec:latency}) while degrading the metric that governs deployment is a poor trade, and PR-AUC gives no warning, improving for exactly the models that fail operationally. Whether to rerank, and with what, must be decided on a cache-aware metric; for a strong domain retriever the answer may be no reranker at all.

\paragraph{Practical Recommendations.} Evaluate cache models with P-CHR AUC or ORR, not PR-AUC, F1, or ROC-AUC, which measure a ranking quality the threshold never uses and hide where scores actually sit relative to it. Prefer objectives whose scores are naturally spread, contrastive or multi-vector, over binary cross-entropy. BCE concentrates most of its mass at the decision boundary and was the worst deployment choice in our study despite its strong PR-AUC. If BCE is unavoidable, re-normalize its scores over the candidate pool (pool-softmax), which recovers much of the gap; post-hoc temperature or Platt scaling will not, as they cannot change a rank-based deployment metric.

Do not assume a reranker always helps: most rerankers we evaluated lowered deployment precision below the retriever alone, so validate the reranking stage on a cache-aware metric before paying its latency cost, and for a strong domain retriever consider thresholding the retriever score directly. We also recommend tuning the decision threshold on data whose positive rate matches the deployment, since both the structural floor and the precision--utilization frontier shift with the duplicate rate. When comparing models across deployments, prefer ORR, which normalizes out this positive-rate-dependent floor and is therefore more portable than absolute P-CHR AUC. Finally, treat threshold selection as an ongoing operational concern rather than a one-time fix: production query distributions drift, so periodically re-estimate the operating threshold on recent, domain-representative traffic.

Taken together, these practices reframe semantic cache model selection around the property deployment actually depends on. Once we measure and optimize threshold utility rather than ranking, the gap between offline promise and production behavior becomes visible, and closing it becomes possible.

\end{document}